\def\arxivmode{1}
\PassOptionsToPackage{expansion=false}{microtype}
\ifdefined\arxivmode
  \documentclass[manuscript,screen,nonacm]{acmart}
\else
  \documentclass[manuscript,screen,review]{acmart}
\fi

\usepackage{graphicx}
\usepackage{booktabs}
\usepackage{amsmath,amsfonts}
\usepackage{multirow}
\usepackage{algorithm}
\usepackage{algpseudocode}
\usepackage{comment}
\usepackage{url}
\usepackage{xcolor}
\usepackage{tabularx}
\usepackage[T1]{fontenc}
\usepackage{pifont}
\usepackage{todonotes}
\usepackage{xspace}
\usepackage{flafter}

\newcommand{\name}{\textsc{XPhysICS}\xspace}
\xspaceaddexceptions{-'}
\newcommand{\namenospace}{\textsc{XPhysICS}}

\ifdefined\arxivmode
  \setcopyright{none}
\else
  \acmJournal{TOSEM}
\fi

\begin{document}

\author{Sangshin Park}
\email{u1418114@utah.edu}
\affiliation{\institution{University of Utah}\city{Salt Lake City}\state{Utah}\country{USA}}
\author{Jainta Paul}
\email{u1471999@utah.edu}
\affiliation{\institution{University of Utah}\city{Salt Lake City}\state{Utah}\country{USA}}
\author{Lawrence Ponce}
\email{u1384059@utah.edu}
\affiliation{\institution{University of Utah}\city{Salt Lake City}\state{Utah}\country{USA}}
\author{Md Raihan Ahmed}
\email{u1374605@utah.edu}
\affiliation{\institution{University of Utah}\city{Salt Lake City}\state{Utah}\country{USA}}
\author{Mu Zhang}
\email{muzhang@cs.utah.edu}
\affiliation{\institution{University of Utah}\city{Salt Lake City}\state{Utah}\country{USA}}
\author{Luis Garcia}
\email{la.garcia@utah.edu}
\affiliation{\institution{Kahlert School of Computing, University of Utah}\city{Salt Lake City}\state{Utah}\country{USA}}


\title{XPhysICS: Cross-Physical-Domain Threat Grounding for Industrial Control Systems Security}


\begin{abstract}
Industrial control system (ICS) attacks are often documented in
plant-specific terms, although their underlying physical effects may transfer
across systems. Determining such transferability requires more than matching
component names or semantic roles.

We present \namenospace, a methodology for grounding documented
cyber--physical threats onto target systems. \name transforms source evidence
into provenance-linked threat abstractions capturing manipulated roles,
physical consequences, and expected observations. Given fixed abstractions
and machine-readable target contracts, deterministic checks assess grounding
eligibility and produce validation slices for downstream evaluation.

We study 83 threat abstractions across continuous-process and manufacturing
sources. The continuous-process evaluation examines 78 abstractions against
water-treatment, water-distribution, hydropower, and chemical-process targets.
Candidate coverage is compared with literal-name and role-only baselines on a
pinned snapshot, while a bounded assignment audit examines agreement across
three judgment streams. Selected cases undergo controlled simulator-signal
perturbations, and downstream compatibility is evaluated using multiple
analysis styles, including upstream GeCo. A bounded realizability study
further examines three objectives on one target using a paper-derived search
reproduction.

The results highlight the importance of explicit target constraints and
traceable evidence when assessing cross-domain threat grounding, while
distinguishing candidate coverage, grounding acceptance, and downstream
validation.
\end{abstract}

\maketitle
\section{Introduction}
\label{sec:intro}

Industrial control systems (ICS) operate physical processes such as water
treatment, power generation, oil and gas handling, and manufacturing, in
which cyber actions can alter physical behavior. Documented incidents include
logic and setpoint modification~\cite{stuxnet-symantec}, protocol abuse for
equipment reconfiguration~\cite{kozak2023industroyer,slowik2018anatomy}, and
safety-system subversion~\cite{di2018triton}. Water-sector incidents further
indicate that chemical-dosing manipulations can recur across distinct
facilities~\cite{penn-water-treatment-plant-attack,
Florida-water-treatment-plant-attack}. Although the affected plants differ,
the underlying pattern may involve recurring physical roles---pumps, valves,
tanks, interlocks, and their control relationships.

Shared incident reports allow operators to learn from such patterns. For
example, the Electricity Information Sharing and Analysis Center of the North
American Electric Reliability Corporation (NERC) published
a technical account of the 2015 Ukrainian power-grid attack for broader
defensive use~\cite{Ukrainian-Power-Grid-Attack}. Such reports, however, are
expressed in terms of the source plant: its tags, vendor-specific components,
control logic, and topology. A dosing-corruption threat is not applicable to
another plant merely because both plants contain chemical tanks; the target
must provide sufficient documented role, process-stage, dependency, and
observation evidence for explicit checks to be applied. The central question
of this paper is therefore: given threat evidence from one ICS and a
description of another, does the source abstraction satisfy the implemented
checks for that target?

Existing representations address related but distinct problems. MITRE
ATT\&CK for ICS~\cite{mitreics} organizes tactics and techniques, whereas
knowledge graphs and ontologies organize assets, threats, and
relationships~\cite{shen2020data,alanen2022hybrid,
mozzaquatro2016ontology,gill2025representingtimecontinuousbehaviorcyberphysical,
neuhaus2009semantic}. These representations support classification and
semantic correspondence, but label similarity does not establish that a
particular target can support or evaluate the same physical effect.
Controller-level analyses, in contrast, can provide detailed target evidence
when the required code and toolchains are available. What remains missing is
a method that connects documented source threats to the evidence available
for a specific system under test (SUT).

We present \namenospace, a framework for this \emph{target-conditioned
grounding} problem. \name represents source evidence by recording the
manipulated roles, the expected physical consequence, the observations that
the source evidence calls for, and the provenance of each claim. It
represents the target through a contract that describes the available roles,
signals, dependencies, and evaluation rules. With these inputs fixed, the
grounder applies five deterministic implementation checks: component
coverage, coarse role/type compatibility, source-stage coherence, minimum-tag
slice viability, and rule-surface intersection. An accepted result can then
be materialized as a \emph{validation slice} that records the target roles,
signals, dependencies, timing assumptions, and provenance selected for
subsequent analysis. Source abstraction may require analyst judgment; the
grounding decision is deterministic only after the abstraction, vocabulary,
schema version, and target contract have been fixed
(Section~\ref{sec:workflow}).

The core procedure does not require target controller code. When such code is
available, \textit{CrossPLC} can contribute routines, tags, read/write
relationships, and state information for role recovery and slice
construction. Its original translation and consolidation path was developed
as part of SCADMAN~\cite{scadman} and is augmented here for \name. This
enrichment, like the temporal guardrails, is optional.

The validation slice also separates grounding from subsequent target-side
analysis. Predictive, state-aware, phase-aware, control-integrity, and
provenance systems~\cite{wolsing2025geco,abbas2024sain,ike2022scaphy,scadman,
ahmed2025icstracker} can consume different views of the target context without
becoming part of the grounding definition. We evaluate this interface using
\name-native consumer lanes, prototype adapters, and the unmodified upstream
GeCo implementation; these are compatibility studies, not comparative
detector evaluations. Dynamic realizability is examined separately by
applying a paper-derived reproduction of the ICSFlux search
method~\cite{11573458} to three frozen groundings.

The evaluation uses source and target settings spanning water treatment,
water distribution, hydropower, chemical processes, and a smaller
manufacturing corpus. It reports broad grounding coverage; a pinned-snapshot
comparison with two simple candidate-mapping baselines, accompanied by a
bounded assignment audit; detailed water-domain validation slices; nine
bounded-support cases on hydropower (Hydro) and GRFICS chemical-process
targets; downstream compatibility studies; and the
three-objective realizability experiment. Because these studies address
different questions, their denominators and conclusions are reported
separately.

The evaluation is organized around four research questions:

\begin{description}
\item[\textbf{RQ1: Structural grounding.}]
Under the evaluated target contracts, which source abstractions satisfy the
implemented grounding predicates across the target process families?

\item[\textbf{RQ2: Slice evidence.}]
For selected accepted groundings, do the constructed validation slices provide
the declared signal, dependency, and timing evidence, and what target-side
consumer outcomes arise under the evaluated perturbations?

\item[\textbf{RQ3: Interoperability.}]
Can the same validation-slice representation supply the inputs required by
distinct downstream analysis styles, including the unmodified upstream GeCo
implementation?

\item[\textbf{RQ4: Dynamic realizability.}]
For a bounded set of frozen groundings, does realizability differ across the
evaluated plant models and command surfaces while structural grounding remains
fixed?
\end{description}

\noindent\textbf{Contributions.}
The primary contribution of this work is a provenance-aware,
target-conditioned methodology for grounding documented ICS threats into
target-specific validation slices. Supporting contributions provide the
representations, mechanisms, and empirical evidence required to evaluate this
methodology.

\begin{itemize}

\item \textbf{Target-conditioned threat grounding.}
We introduce a framework that separates analyst-guided abstraction of
source-side threat evidence from deterministic target grounding. Once the
source abstraction, vocabulary and schema version, and target contract are
fixed, candidate mappings are evaluated using five explicit eligibility
criteria covering component coverage, implemented role/type compatibility,
source-stage coherence, minimum-tag slice viability, and rule-surface
intersection
(Sections~\ref{sec:framework:extraction}--\ref{sec:framework:grounding}).
The resulting grounding record distinguishes candidate-generation outcomes
from rejection under individual eligibility criteria.

\item \textbf{Validation slices for target-side evaluation.}
We introduce the validation slice as a target-specific representation that
records mapped manipulation and consequence paths, observable signals,
dependencies, timing assumptions, and consumer-relevant context
(Section~\ref{sec:framework:slice}). This representation enables grounding
acceptance, slice adequacy, consumer applicability, and consumer outcome to be
evaluated separately rather than represented by a single cross-system
applicability decision.

\item \textbf{Machine-validated evidence and target-contract representations.}
We provide schema-validated representations for source abstractions and target
contracts (Table~\ref{tab:semantic-constructs} and
Table~\ref{tab:target-contract-fields}), together with versioned vocabularies
and an evidence ledger that preserves provenance across source evidence,
abstraction fields, and grounding decisions. Optional CrossPLC-based
controller enrichment, temporal guardrails, and bounded downstream consumers
evaluate additional uses of the resulting validation-slice representation.

\item \textbf{Multi-domain evaluation with explicit evidence levels.}
We evaluate \name across water-treatment, water-distribution,
hydro/water-energy, and chemical-process settings. The evaluation reports
continuous-process grounding coverage, detailed executable water-domain
validation slices, nine bounded-support Hydro/GRFICS cases, downstream
consumer compatibility, and a bounded dynamic-realizability study. These
results characterize distinct evidence levels and identify the scope within
which cross-domain threat semantics remain structurally groundable and
target-side evaluable.

\end{itemize}

\section{Operational Context and Problem Statement}
\label{sec:context}

\begin{figure*}[t]
  \centering
  \includegraphics[width=0.78\linewidth,trim=0 8 0 8,clip]{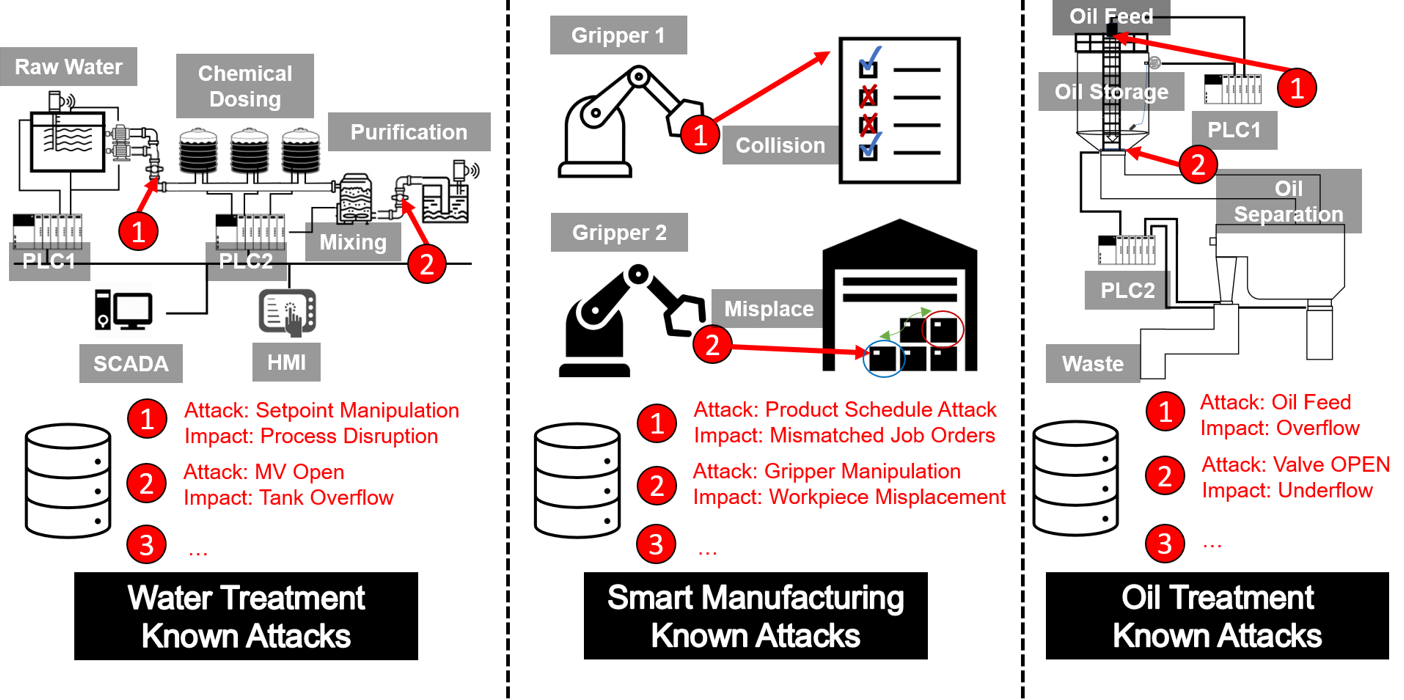}
  \caption{Recurring cyber--physical threat patterns across ICS domains.
  Water-treatment, manufacturing, and oil/chemical-process systems may expose
  similar physical roles, including pumps, valves, tanks, interlocks, and
  control interactions, despite differences in plant topology, process
  structure, and physical medium.}
  \label{fig:usecase}
\end{figure*}

\noindent\textbf{Why plant-specific grounding matters.}
Figure~\ref{fig:usecase} illustrates the distinction between recurring
cyber--physical roles and target-specific applicability. Components such as
pumps, valves, tanks, and interlocks may appear across multiple plants without
providing equivalent physical roles, process context, or observability. For
example, an overflow-related manipulation documented for one water-treatment
stage may have candidate counterparts on another target, but component-name
or type similarity alone does not establish whether the target can evaluate
the corresponding effect. The implemented grounding filters check component
coverage, role/type, source-stage, minimum-tag slice viability, and rule
intersection; complete source-role coverage and slice adequacy require
separate evidence.

\subsection{Problem Statement}
\label{sec:problem}

Given source artifacts describing an ICS threat and target artifacts describing
a selected SUT, \name determines whether the source threat abstraction is
structurally admissible under the target contract. For an accepted grounding,
\name constructs a target-specific \emph{validation slice} containing the
manipulated and consequence paths, observable signals, dependencies, timing
assumptions, and consumer-relevant context required for subsequent target-side
evaluation. A \emph{rule surface} denotes the target variables, dependencies,
and operating context over which the target contract defines applicable
evaluation rules.

The problem therefore comprises three stages: constructing a reusable threat
abstraction from heterogeneous source evidence, evaluating candidate mappings
against explicit target-conditioned eligibility criteria, and constructing a
validation slice whose adequacy can be assessed independently of the grounding
decision. Downstream consumer applicability and outcome are evaluated
subsequently and do not determine grounding acceptance.
Section~\ref{sec:framework} formalizes these stages.

\begin{figure*}[t]
  \centering
  \includegraphics[width=\linewidth]{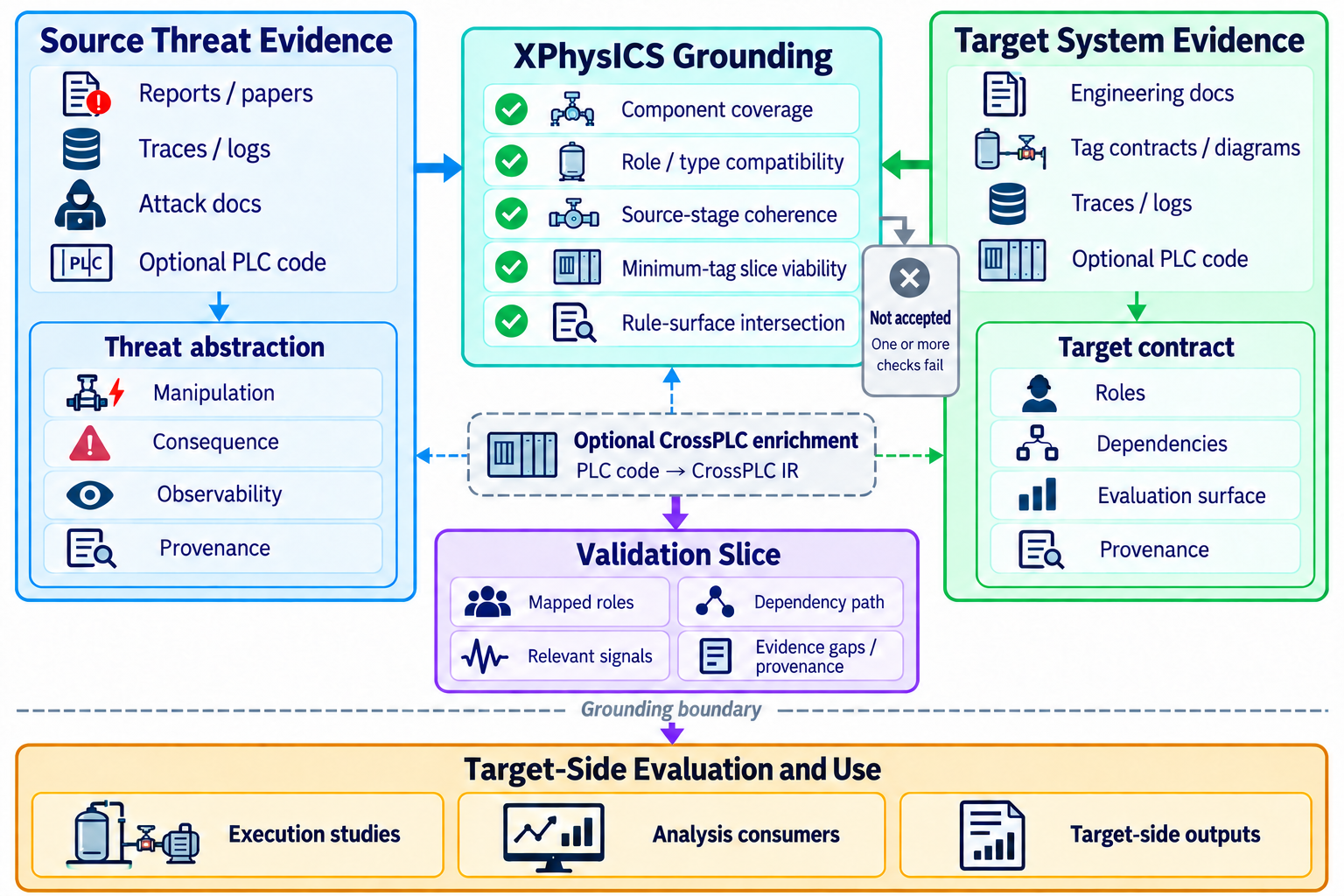}
  \caption{Overview of \name. Evidence from a documented source threat and a
  target system is converted into a threat abstraction and target contract.
  The grounder applies the implemented eligibility checks and returns a
  target-specific validation slice when all checks pass; otherwise, the
  candidate is not accepted.
  Execution and downstream analysis occur beyond the grounding boundary.
  CrossPLC provides optional controller-code enrichment.}
  \label{fig:framework}
\end{figure*}

\subsection{Use by Downstream Analyses}
\label{sec:ops-touchpoints}

Validation slices provide a common target-conditioned representation for
downstream analyses with different input requirements. Invariant-based
analyses may consume ranges, correlations, mass-balance relationships, or
rate-of-change constraints; predictive analyses may require time-series
windows and expected trajectories; state- or phase-aware analyses additionally
require operating context; and provenance-oriented analyses may use links
between target-side observations, grounding decisions, and their supporting
evidence. The validation slice exposes the target roles, signals,
dependencies, contextual information, and provenance needed to construct
these consumer-specific views.

Consumer applicability remains separate from grounding and slice adequacy.
A validation slice may be structurally grounded and adequate for target-side
evaluation while falling outside the modeled surface of a particular
consumer. Similarly, the outcome produced by an applicable consumer does not
alter the underlying grounding decision. This separation allows the same
grounded representation to support heterogeneous downstream analyses without
making any individual consumer part of the definition of grounding.

\subsection{Design Requirements and Scope}
\label{sec:reqs}

The operational setting motivates four design requirements.
\textbf{Target specificity} requires each grounding decision to be conditioned
on a specified SUT and its available evidence rather than represented as an
abstract semantic correspondence.
\textbf{Explicit eligibility} requires candidate mappings to satisfy the
declared implementation-level filters before grounding acceptance, while
recording limitations those filters do not establish.
\textbf{Slice adequacy} requires the resulting validation slice to expose the
signal coverage, dependency closure, and timing evaluability needed for the
intended target-side evaluation, with adequacy reported separately from
grounding and consumer outcome.
\textbf{Auditable provenance} requires source evidence, target evidence,
mapping decisions, and eligibility outcomes to remain traceable throughout
the grounding process. Section~\ref{sec:framework:grounding} formalizes the
grounding criteria, while Section~\ref{sec:framework:adequacy} defines the
subsequent adequacy assessment.

\name operates over heterogeneous ICS artifacts, including attack
descriptions, dataset documentation, engineering diagrams, manuals, target
contracts, time-series traces, and controller artifacts when available.
Controller artifacts are not required by the core grounding procedure; when
present, they provide optional evidence for role recovery and
validation-slice construction through program-side tags, routines, read/write
relationships, and available state information.

The primary evaluation focuses on continuous-process settings spanning water
treatment, water distribution, hydro/water-energy, and chemical processes.
Additional studies examine manufacturing/program-artifact grounding,
controller-code enrichment, temporal guardrails, and protocol evidence under
separate evidence surfaces and denominators. These extension studies are
reported independently of the continuous-process grounding matrix and the
primary validation-slice studies.

\section{\name Framework}
\label{sec:framework}

Figure~\ref{fig:framework} summarizes the \name methodology. The framework
connects source-side threat evidence to target-conditioned security
evaluation through four stages: source threat abstraction, target-conditioned
grounding, validation-slice construction, and slice adequacy with downstream
interfaces. The objective is not to establish equivalence between source and
target plants, but to determine whether documented source threat semantics
satisfy explicit structural and evaluability requirements for a specified
target SUT.

We represent a source evidence bundle as
\[
\mathcal{S} = (\mathcal{D}_s, \mathcal{X}_s, \mathcal{P}_s),
\]
where $\mathcal{D}_s$ denotes source documents and engineering artifacts,
$\mathcal{X}_s$ denotes traces or benchmark data when available, and
$\mathcal{P}_s$ denotes optional controller or program artifacts. A target SUT
is represented as
\[
\mathcal{T} = (\mathcal{D}_t, \mathcal{X}_t, \mathcal{P}_t, \mathcal{R}_t),
\]
where $\mathcal{R}_t$ denotes target-side rule surfaces or consumer-relevant
variables. Rather than representing a comprehensive plant model, \name
produces a grounded, target-specific validation object that provides
structured input for subsequent target-side evaluation.

\begin{figure}[t]
  \centering
  \includegraphics[width=\linewidth]{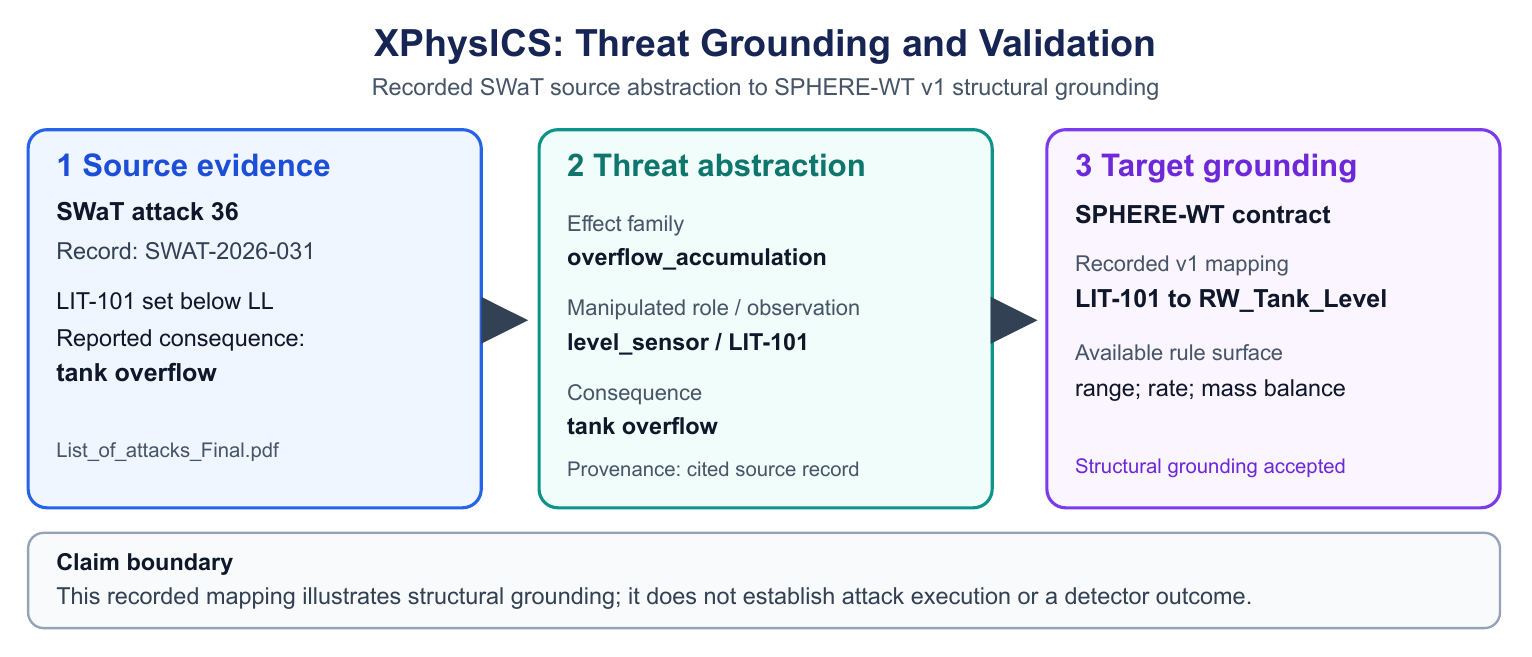}
  \caption{Source-backed worked grounding example. Secure Water Treatment
  (SWaT) attack 36 is recorded as \texttt{SWAT-2026-031}, with a LIT-101
  low-reading manipulation and a reported tank-overflow consequence. The
  recorded v1 structural grounding maps its level-sensor role to the
  \texttt{RW\_Tank\_Level} signal of the SPHERE water-treatment (WT) target.}
  \label{fig:semantic-example}
\end{figure}

\noindent\textbf{Running example.}
Figure~\ref{fig:semantic-example} illustrates the principal transformations
performed by \name. Source evidence describing a cyber--physical manipulation
is first represented in terms of its effect family, manipulated roles,
consequence roles, observability obligations, and provenance. Grounding then
attempts to map these source-side requirements onto target entities and
signals and evaluates the resulting candidate against the target contract.
When the grounding criteria are satisfied, the mapped entities and signals
form the basis of a target-specific validation slice.

This example illustrates the distinction between semantic similarity and
target-conditioned grounding. A semantically related tank is insufficient to
establish an admissible grounding. Grounding is accepted only when the candidate
satisfies the implemented component-coverage, role/type, source-stage,
minimum-tag, and rule-intersection filters. The example is a recorded structural
mapping, not an executed attack or a measured consumer outcome.

\subsection{Source Threat Abstraction}
\label{sec:framework:extraction}

The first transformation maps heterogeneous source evidence into a structured
threat abstraction. Source evidence may include engineering documentation,
manuals, incident reports, benchmark traces, attack annotations, and, where
available, controller or program artifacts. \name represents these
heterogeneous evidence sources within a common abstraction schema, providing
the structured source semantics required for subsequent target-conditioned
grounding.

For each candidate source-side threat pattern, \name constructs
\[
\tau = (f, R_m, R_c, O, \Pi),
\]
where $f$ is the \emph{effect family}, $R_m$ is the set of
\emph{manipulated roles}, $R_c$ is the set of \emph{consequence roles},
$O$ is the set of \emph{observability obligations}, and $\Pi$ is the
\emph{provenance bundle} linking the abstraction to its supporting source
evidence. This representation preserves the manipulated roles,
expected physical consequence, observations required to assess that
consequence, and supporting evidence required for subsequent grounding.

Table~\ref{tab:semantic-constructs} provides an operational definition of each
field, including its representation and data type, the vocabulary constraining
its admissible values, the supporting source evidence, and the mechanism by
which the field is populated, whether through analyst interpretation,
source-specific parsing, or inheritance from prior work. The provenance bundle
$\Pi$ maintains traceability between each populated field value and its
supporting source evidence.

\begin{table}[t]
\centering
\small
\caption{Operational definitions of the source threat abstraction
$\tau=(f,R_m,R_c,O,\Pi)$. Origin distinguishes OTThreat-informed constructs
from \name-specific constructs.}
\label{tab:semantic-constructs}
\begin{tabularx}{\linewidth}{@{}p{0.11\linewidth}Xp{0.14\linewidth}@{}}
\toprule
\textbf{Construct} & \textbf{Operational definition} &
\textbf{Origin / assignment} \\
\midrule
Effect family $f$ &
Normalized physical-consequence category drawn from a closed, versioned
extraction vocabulary. Target grounding operates over the narrower grounding
vocabulary declared in scope by the target contract
(Section~\ref{sec:framework:grounding}). &
OTThreat-informed vocabulary; analyst- or parser-assigned. \\
\addlinespace

Manipulated roles $R_m$ &
Source-side component roles manipulated by the threat, represented by an
identifier and functional role with optional process-stage context. &
\name-specific; analyst- or parser-assigned. \\
\addlinespace

Consequence roles $R_c$ &
Components or consequence records associated with the realized physical
effect $f$ following manipulation of $R_m$. &
\name-specific; assigned from supporting source evidence. \\
\addlinespace

Observability obligations $O$ &
Signals or observations indicated by the source evidence as relevant to
assessing the stated effect; these obligations seed target-side slice
construction (Section~\ref{sec:framework:slice}). &
\name-specific; may be empty when the source evidence does not identify
corresponding observations. \\
\addlinespace

Provenance $\Pi$ &
Evidence-lineage metadata linking populated abstraction fields to the source
document, supporting excerpt, extraction method, and confidence assignment. &
OTThreat-informed; provenance metadata are recorded by the evidence workflow,
while confidence may involve analyst or reviewer judgment. \\
\bottomrule
\end{tabularx}
\end{table}

\noindent\textbf{Worked example.}
For the source record in Figure~\ref{fig:semantic-example}, the analyst
assigns $f=\texttt{overflow\_accumulation}$ based on the explicitly reported
tank-overflow consequence. $R_m$ contains \texttt{LIT-101}
(role \texttt{level\_sensor}), the named manipulated signal in Secure Water
Treatment (SWaT) attack 36.
The source-evidenced observability set is $O=\{\texttt{LIT-101}\}$; no pump,
valve, or flow tag is added to this source record merely because it might be
useful downstream. The provenance bundle $\Pi$ records
\texttt{List\_of\_attacks\_Final.pdf}, attack 36, the
\texttt{structured\_parse} extraction method, and \texttt{high} confidence.
The v1 grounder records an accepted structural mapping of this abstraction to
the \texttt{RW\_Tank\_Level} signal of the SPHERE water-treatment (WT) target.
The target contract
offers range, rate, and mass-balance rule surfaces, but their availability
does not itself establish a detector outcome for this source record.

Construction of $\tau$ may require analyst interpretation or source-specific
parsing when the available evidence is ambiguous. For example, an observation
such as ``LIT-101 reads erratically'' without an explicit consequence does not
uniquely determine $f$; such evidence requires additional interpretation or an
appropriately qualified confidence assignment. This semantic-abstraction step
is distinct from target grounding. Once $\tau$, the vocabulary and schema
version, and the target contract are fixed, the grounding procedure in
Section~\ref{sec:framework:grounding} applies the specified eligibility
criteria deterministically without additional semantic interpretation.

When programmable logic controller (PLC) artifacts or program artifacts are
available, \textit{CrossPLC} provides optional controller-aware enrichment of
the source or target evidence layer.
We represent this enrichment as
\[
\mathcal{I} = (Q, T, E^{rw}, E^{call}, S),
\]
where $Q$ denotes routines or function blocks, $T$ denotes controller tags,
$E^{rw}$ represents read/write dependencies, $E^{call}$ represents call
relations, and $S$ denotes recovered controller-state structure when
available. This information can provide additional evidence for role recovery
and validation-slice construction but is not required by the core grounding
procedure.

\subsection{Analyst Workflow}
\label{sec:workflow}

The \name workflow separates analyst-guided source abstraction from
deterministic target grounding and subsequent target-side evaluation:

\begin{enumerate}
    \item \textbf{Source evidence} (input). The workflow begins with source
    artifacts, such as engineering documentation, dataset descriptions,
    incident reports, or structured workbooks, that describe a source-side
    threat.

    \item \textbf{Evidence selection} (manual/parser-assisted). An analyst or
    source-specific parser (Section~\ref{sec:impl-source-evidence}) identifies
    evidence describing the manipulated components, physical consequence, and
    relevant observations.

    \item \textbf{Semantic abstraction} (manual/parser-assisted). The selected
    evidence is represented as $\tau=(f,R_m,R_c,O,\Pi)$ using the vocabulary
    defined in Table~\ref{tab:semantic-constructs}. Unstructured evidence may
    require analyst interpretation, whereas compatible structured artifacts
    may be processed by source-specific parsers.

    \item \textbf{Normalization} (automated). Component identifiers and role
    labels are canonicalized according to the schema-defined vocabulary.
    Values outside the permitted enumerations are rejected.

    \item \textbf{Source-evidence traceability} (manual). Source locations and
    excerpts are retained with the structured records so that their
    provenance can be inspected. This step does not independently adjudicate
    abstraction semantics or target mappings.

    \item \textbf{Target-contract construction} (manual, per target). An
    engineer specifies the available target signals, dependency relationships,
    effect-family scope, rule surfaces, and relevant study context in a target
    contract (Table~\ref{tab:target-contract-fields}). The contract is defined
    at the target-SUT level rather than independently for each source threat.

    \item \textbf{Candidate mapping} (automated). The grounder constructs
    candidate mappings from $R_m\cup R_c\cup O$ to entities and signals
    declared in the target contract (Algorithm~\ref{alg:grounding}).

    \item \textbf{Eligibility evaluation} (automated). The predicates
    $\chi_{\text{map}}, \chi_{\text{type}}, \chi_{\text{stage}},
    \chi_{\text{slice}}, \chi_{\text{rule}}$ are evaluated for the fixed
    source abstraction and target contract without additional semantic
    interpretation.

    \item \textbf{Validation-slice construction} (automated). An accepted
    grounding is materialized as
    $\mathcal{V}(\tau,\mathcal{T},\mu)$.

    \item \textbf{Downstream evaluation} (optional). Slice adequacy and
    consumer applicability are assessed independently of grounding acceptance,
    after which applicable consumers may be evaluated over the resulting
    slice.
\end{enumerate}

Open-ended semantic interpretation is concentrated in source evidence
selection and abstraction, whereas normalization, candidate mapping,
eligibility evaluation, validation-slice construction, and downstream
consumer execution operate according to predefined procedures once their
inputs are fixed. Target-contract construction and bounded review additionally
require manual input. The workflow therefore distinguishes analyst-guided
semantic abstraction from deterministic target grounding without
characterizing the complete process as either fully automated or fully manual.

\subsection{Target-Conditioned Grounding}
\label{sec:framework:grounding}

Given a source abstraction $\tau$, \name attempts to instantiate it on a
specified target SUT. The target provides a grounding surface
\[
\Gamma_t = (V_t, E_t, U_t, \Sigma_t, \mathcal{R}_t),
\]
where $V_t$ denotes target entities or process roles, $E_t$ denotes dependency
or flow relationships, $U_t$ denotes unit and engineering metadata,
$\Sigma_t$ denotes observable target signals, and $\mathcal{R}_t$ denotes
the target rule surface. When controller-aware enrichment is available, the
grounding surface may additionally incorporate $\mathcal{I}_t$.

A candidate grounding is a partial mapping
\[
\mu : R_m \cup R_c \cup O \rightharpoonup V_t \cup \Sigma_t
\]
from source-side roles and observability obligations to target-side entities
and signals. The v1 implementation permits individual source components or
observation obligations to remain unresolved; its component-coverage
predicate requires at least one mapped component. \name accepts the resulting
candidate only when
\[
G(\tau,\mathcal{T},\mu) =
\chi_{\text{map}}(\mu)\wedge
\chi_{\text{type}}(\mu)\wedge
\chi_{\text{stage}}(\mu)\wedge
\chi_{\text{slice}}(\mu)\wedge
\chi_{\text{rule}}(\mu).
\]

Here, $\chi_{\text{map}}$ records whether at least one source component has a
candidate target mapping. $\chi_{\text{type}}$ applies the implemented coarse
source-role/target-tag-type check to each mapped component; it is not a
physical-unit checker. $\chi_{\text{stage}}$ checks whether numeric
process-stage labels declared on source components span at most adjacent
stages. $\chi_{\text{slice}}$ expands mapped target tags through declared
target relationships and requires the resulting tag set to meet a
target-specific minimum size. Finally, $\chi_{\text{rule}}$ requires at least
one target rule in $\mathcal{R}_t$ to reference a tag in that expanded set
(Section~\ref{sec:impl-target}). These are implementation-level admissibility
filters. Grounding acceptance does not establish complete role coverage, a
target-side manipulation-to-consequence path, or validation-slice adequacy;
those stronger properties require separate evidence.

The extraction and grounding vocabularies have distinct scopes. Source
abstractions may use the seven effect families defined in
Table~\ref{tab:semantic-constructs}, whereas the reported grounding evaluation
is restricted to a five-family subset. Abstractions assigned to either of the
two extraction-only families therefore fall outside the reported grounding
surface rather than constituting rejected source threats. This distinction
preserves the broader source representation while delimiting the effect
families covered by the current grounding results.

Given a fixed source abstraction, vocabulary and schema version, and target
contract, the grounding procedure evaluates these eligibility criteria
deterministically. The resulting decision establishes whether the candidate
satisfies the implemented component-coverage, role/type, source-stage,
minimum-tag, and rule-intersection filters. It does not establish that every
source role was resolved or that a target dependency path connects the
manipulation and consequence roles. Controller-aware evidence may supplement
subsequent slice construction when program artifacts are available but is not
required by the core grounding procedure.

\noindent\textbf{Grounding and downstream evaluation are distinct.}
The $\chi_{\text{rule}}$ predicate establishes that an accepted grounding
intersects at least one evaluation rule declared by the target contract.
Grounding acceptance does not depend on the applicability or outcome of the
bounded predictive, state-aware, or phase-aware consumers, nor on that of the
upstream GeCo implementation evaluated in
Section~\ref{sec:eval-geco-real}. Consumer applicability is assessed only
after grounding acceptance and validation-slice construction.

Accordingly, \name distinguishes multiple evidence layers.
\emph{Grounding acceptance} establishes acceptance under the implemented
structural-admissibility filters for the target contract.
\emph{Slice adequacy} (Section~\ref{sec:framework:adequacy}) assesses signal
coverage, dependency closure, and timing evaluability.
\emph{Dynamic realizability}, when evaluated, asks whether the grounded effect
can be produced under a specified plant model and command surface
(Section~\ref{sec:eval-realizability}).
\emph{Consumer applicability} determines whether a particular downstream
analysis can operate on the resulting slice.
Finally, \emph{consumer outcome} records the result produced by an applicable
consumer.

\subsection{Validation Slice Construction}
\label{sec:framework:slice}

For an accepted grounding, \name instantiates a target-specific validation
slice
\[
\mathcal{V}(\tau,\mathcal{T},\mu) = (P_m, P_c, \Sigma_v, \Omega_v),
\]
where $P_m$ is the \emph{manipulated path}, $P_c$ is the
\emph{consequence path}, $\Sigma_v \subseteq \Sigma_t$ is the set of
slice-observable signals, and $\Omega_v$ is the \emph{consumer obligation
set}.

The validation slice is the core methodological object of \name. It defines
the bounded, target-specific structure that preserves the manipulated control
surface, expected effect path, observable signals required to assess the
effect, and obligations required by subsequent target-side analyses. This
representation provides an intermediate abstraction between source-side
semantic correspondence and whole-plant equivalence. Semantic correspondence
alone does not establish that a target exposes sufficient structure for
evaluation, whereas requiring equivalence between complete plants would impose
unnecessary constraints on target-specific analysis. The validation slice
therefore captures the target-side structure required to evaluate a grounded
threat under the available evidence.

\subsection{Slice Adequacy and Downstream Interfaces}
\label{sec:framework:adequacy}

The adequacy of a grounded slice is assessed with respect to the requirements
of the intended target-side evaluation. \name characterizes slice adequacy
along three dimensions: signal coverage, dependency closure, and timing
evaluability. These criteria assess whether the slice provides the
observations, process relationships, and temporal resolution required to
evaluate the grounded effect.

Slice adequacy is classified as \emph{strong}, \emph{partial}, or
\emph{insufficient}. A \emph{strong} slice satisfies the declared adequacy
requirements for the intended evaluation. A \emph{partial} slice retains an
accepted grounding but has identified limitations in signal coverage,
dependency closure, or timing evaluability. An \emph{insufficient} slice does
not satisfy the minimum adequacy requirements for the intended target-side
evaluation. For strong or partial slices, \name may derive consumer-specific
views from the grounded slice and its associated provenance. These views
support invariant-family analyses, bounded predictive consumers, bounded
state- or phase-aware consumers, provenance-oriented analyses, and optional
temporal guardrails.

Adequacy labels are distinct from consumer applicability and outcome labels.
A strong slice may, for example, be applicable to a particular consumer yet
produce a \emph{nominal-confounded} or \emph{near-threshold} outcome.
Conversely, a grounded slice may fall outside the modeled surface of a
particular consumer without changing its grounding or adequacy classification.
These distinctions preserve the evidentiary scope of each stage throughout
the evaluation in Section~\ref{sec:eval}.

\section{Implementation}
\label{sec:implementation}

We implement \name as a target-conditioned pipeline that transforms
source-side threat evidence and target-side system evidence into structured
validation artifacts for a selected SUT. Following
Figure~\ref{fig:framework}, source artifacts are represented as threat
abstractions, target evidence is specified through machine-validated
contracts, deterministic grounding evaluates candidate mappings, and accepted
groundings are materialized as validation-slice bundles for subsequent
target-side evaluation.

\subsection{Evidence Records and Threat Abstractions}
\label{sec:impl-source-evidence}

\name begins by registering source artifacts that describe reusable ICS threat
patterns, including incident reports, academic attack papers, benchmark
annotations, engineering documentation, structured workbooks, and selected
controller artifacts. Each artifact is represented by a source-document record
containing its domain, document identifier, provenance tier, admissibility
metadata, source location, and extraction metadata. These records preserve the
provenance required to relate structured threat abstractions to their
supporting source evidence.

The extractor converts source records into \texttt{ThreatAbstraction} objects
implementing the representation defined in
Section~\ref{sec:framework:extraction}. Each object records the effect family,
manipulated roles, consequence roles, observability obligations, provenance
links, mechanism metadata, timing information when available, confidence,
supporting source evidence, and extraction metadata. The primary evaluation
uses structured or curated extraction paths. Extraction assisted by large
language models (LLMs) is retained as an exploratory support mechanism and is
not used as the basis for the reported grounding results.

For sources that provide structured local artifacts, \name employs
source-specific parsers to construct the common threat abstraction. The Water
Distribution (WADI) parser, for example, extracts attack points, actuator and
sensor state changes, physical impacts, and labeled attack intervals from
workbook and trace artifacts, supplementing evidence available in
document-based sources. Critical Infrastructure Security Showdown (CISS) data
and other supporting sources are processed analogously when compatible
structured artifacts are available. Regardless of the source-specific
extraction procedure, each parser produces the common abstraction schema
required by the subsequent grounding procedure.

\subsection{Target Contracts and Study Bundles}
\label{sec:impl-target}

Grounding is conditioned on a selected target SUT. We represent the target
grounding surface $\Gamma_t$ (Section~\ref{sec:framework:grounding}) using
machine-validated target contracts that describe available target signals and
roles, dependency relationships, effect-family scope, and the rules and
operating context available for target-side evaluation. Study bundles
associate these contracts with nominal traces, controlled perturbation cases,
and timing information used to exercise grounded validation slices.

The primary target studies use contracts for the SPHERE~\cite{garcia2025sphere}
WT and water-distribution (WD) systems. Additional hydropower, chemical-process, CISS~\cite{itrust2019ciss}, and manufacturing
studies use the same representation where the required target evidence is
available. Their results are reported under the corresponding evidence scope
and denominator in Section~\ref{sec:eval}.

Target contracts are schema-validated before grounding or execution.
Contracts that do not satisfy the required schema are rejected before
grounding rather than interpreted with missing fields.
Table~\ref{tab:target-contract-fields} summarizes the components most relevant
to grounding and validation-slice construction. The reported experiments use
versioned schemas and executable study contracts.

\begin{table}[t]
\centering
\small
\caption{Target-contract components used by \name for grounding and
validation-slice construction.}
\label{tab:target-contract-fields}
\begin{tabularx}{\linewidth}{@{}p{0.28\linewidth}X@{}}
\toprule
\textbf{Contract component} & \textbf{Role in \name} \\
\midrule

Effect-family scope &
Declares the source-threat effect families within the target's current
grounding scope. \\

Target signals and roles &
Defines the target signal-and-role surface over which candidate mappings are
evaluated. \\

Dependency relationships &
Provides target relationships used during validation-slice construction and
dependency/connectivity evaluation. \\

Threat-surface scope &
Declares the boundary directions and threat classes represented by the
target study. \\

Target rules and context &
Defines the target rules and operating context used to determine rule-surface
applicability and support subsequent slice evaluation. \\

\bottomrule
\end{tabularx}
\end{table}

The target-contract grounding vocabulary is a five-family subset of the
seven-family extraction vocabulary introduced in
Table~\ref{tab:semantic-constructs}. Consequently, a source abstraction may
be valid within the extraction representation while falling outside the
effect-family scope supported by the current target contracts. Such
abstractions are reported outside the current grounding surface rather than as
rejected source threats.

\subsection{Deterministic Grounding and Slice Generation}
\label{sec:impl-grounding}

The deterministic grounder evaluates each source abstraction $\tau$ against a
target grounding surface
$\Gamma_t=(V_t,E_t,U_t,\Sigma_t,\mathcal{R}_t)$, where $V_t$ denotes target
entities or roles, $E_t$ denotes dependency or flow relationships, $U_t$
denotes unit and engineering metadata, $\Sigma_t$ denotes observable target
signals, and $\mathcal{R}_t$ denotes the target rule surface. Candidate
mappings from source roles and observability obligations to target entities
and signals are evaluated using the five eligibility predicates defined in
Section~\ref{sec:framework:grounding}.

The resulting grounding record contains the source threat identifier, target
SUT, mapped tags or components, applicable target rules, slice requirements,
per-check diagnostics, and provenance links. Grounding acceptance establishes
acceptance under the implemented structural-admissibility filters for the
target contract.

Algorithm~\ref{alg:grounding} specifies the grounding procedure at the
granularity implemented by \name. Candidate construction applies an ordered
matching procedure consisting of exact or normalized tag-name matching,
configured aliases, role-preference lookup, and a role-keyword search over
target-tag descriptions. The first compatible candidate in the declared
contract order is selected, and the selected mapping method and associated
metadata are retained in the grounding record. Determinism here follows from
fixed inputs, fixed precedence, and fixed contract order.

The implemented role/type check performs a coarse compatibility test between
the source role (e.g., \texttt{level\_sensor}) and the declared target-tag
type (e.g., \texttt{real} or \texttt{bool}). Stage coherence uses numeric process-stage labels
declared on source components and excludes candidates spanning non-adjacent
source stages; it does not inspect the stages of mapped target tags.
Observability obligations are resolved separately by exact, normalized, or
configured-alias lookup, and unresolved obligations are omitted from the
initial expansion set. Slice viability expands the mapped target-tag set
through the correlation, causality, mass-balance, and trip-response
relationships declared by the target contract until no additional tags are
introduced and evaluates the resulting set against the target-specific
minimum slice size. Rule applicability requires at least one declared target
rule to intersect the resulting tag set.

The current component-coverage predicate requires at least one source
component mapping, rather than total coverage of every component or
observation obligation. Grounding acceptance requires all five eligibility
predicates to hold. For
accepted and rejected candidates, the grounding record retains the outcome of
each predicate, providing an explicit basis for the resulting decision.

\begin{algorithm}[t]
\caption{Deterministic grounding of one source abstraction $\tau$ against one
target contract $\Gamma_t$.}
\label{alg:grounding}
\begin{algorithmic}[1]
\Require source abstraction $\tau=(f,R_m,R_c,O,\Pi)$; target contract
$\Gamma_t=(V_t,E_t,U_t,\Sigma_t,\mathcal{R}_t)$
\Ensure grounding record with success flag, tag mappings, applicable rules,
slice requirements, and per-check diagnostics

\State $M \gets \emptyset$
\Comment{candidate tag mappings}

\For{each component $c \in R_m \cup R_c$}
    \State $m \gets$ \Call{MapComponentToTag}{$c$, $\Gamma_t$}
    \Comment{exact $\to$ normalized $\to$ alias $\to$ role-preference $\to$ lexical}
    \If{$m \neq \bot$}
        $M \gets M \cup \{m\}$
    \EndIf
\EndFor

\State $\chi_{\text{map}} \gets (M \neq \emptyset)$

\State $\chi_{\text{type}} \gets
\chi_{\text{map}} \wedge
\bigwedge_{m \in M}$ \Call{RoleTypeCompatible}{$m$}

\State $S \gets
\{\,\text{process\_stage}(c) : c \in R_m \cup R_c\,\}$

\State $\chi_{\text{stage}} \gets$
(stages in $S$ are pairwise adjacent, or $|S|\le 1$)

\State $O_t \gets \emptyset$
\For{each observation obligation $o \in O$}
    \State $s \gets$ \Call{FindTargetTag}{$o$, $\Gamma_t$}
    \Comment{exact $\to$ normalized $\to$ configured alias}
    \If{$s \neq \bot$}
        $O_t \gets O_t \cup \{s\}$
    \EndIf
\EndFor

\State $T \gets
\Call{ExpandToFixpoint}{M \cup O_t, \mathcal{R}_t}$
\Comment{correlation/causality/mass-balance/trip-response relationships}

\State $\chi_{\text{slice}} \gets
(|T| \ge \text{minimum\_slice\_tags})$

\State $A \gets
\{r \in \mathcal{R}_t :
\text{tags}(r) \cap T \neq \emptyset\}$

\State $\chi_{\text{rule}} \gets (A \neq \emptyset)$

\State $\textit{success} \gets
\chi_{\text{map}} \wedge
\chi_{\text{type}} \wedge
\chi_{\text{stage}} \wedge
\chi_{\text{slice}} \wedge
\chi_{\text{rule}}$

\State \Return
$(\textit{success},M,A,T,
(\chi_{\text{map}},\chi_{\text{type}},\chi_{\text{stage}},
\chi_{\text{slice}},\chi_{\text{rule}}))$

\end{algorithmic}
\end{algorithm}

Accepted groundings are materialized as validation-slice bundles implementing
$\mathcal{V}(\tau,\mathcal{T},\mu)=(P_m,P_c,\Sigma_v,\Omega_v)$ as defined in
Section~\ref{sec:framework:slice}. Each bundle records the manipulated and
consequence paths, required signals and dependencies, perturbation
specifications when applicable, nominal and perturbed traces when available,
timing assumptions, operating context, adequacy metadata, and consumer-facing
outputs. Rather than representing a complete digital twin or exhaustive model
of the target, the bundle encodes the bounded target-specific structure
required for subsequent evaluation of the grounded effect. Slice adequacy is
assessed independently in terms of signal coverage, dependency closure, and
timing evaluability, yielding \emph{strong}, \emph{partial}, or
\emph{insufficient} classifications.

\subsection{Bounded Validation Consumers}
\label{sec:impl-consumers}

Downstream consumers operate on validation slices rather than directly on
source evidence. The implementation includes four bounded consumer families.
The invariant-family consumer evaluates target-declared range,
rate-of-change, correlation, and mass-balance rules over slice signals and
dependencies and serves as the primary native consumer for the WT/WD
validation studies. A GeCo-style predictive lane~\cite{wolsing2025geco}
trains a predictive model, calibrates thresholds using target nominal traces,
and evaluates residual or cumulative-sum deviations over perturbation cases.
A SAIN-style state-aware lane~\cite{abbas2024sain} infers coarse target states
from available tags and applies state-conditioned thresholds to residual
invariants. A SCAPHY-style phase-aware lane~\cite{ike2022scaphy} infers coarse
operating phases from target state and actuator tags and applies
phase-conditioned value/delta scoring with persistence-gated alerts.

The GeCo-, SAIN-, and SCAPHY-style lanes are bounded \name-native
implementations of selected analysis patterns motivated by the corresponding
published systems. They are used to evaluate whether the validation-slice
representation provides the information required by these downstream analysis
styles. Each lane produces consumer-specific measurements, including detected
windows, alert timing, summary metrics, nominal false-positive rates, and
case-level outcomes. These measurements characterize the bounded
implementations rather than comparative performance against the corresponding
published systems. Compatibility with the upstream GeCo implementation is
evaluated separately in Section~\ref{sec:eval-geco-real}.

\subsection{Optional Enrichment Paths}
\label{sec:impl-enrichment}

\name supports two optional enrichment mechanisms: controller-code evidence
and temporal guardrails. These mechanisms supplement particular validation
slices when the required artifacts are available but are not prerequisites for
the core grounding procedure.

When source or target PLC/program artifacts are available, \name can invoke
CrossPLC. Its original controller-code translation and consolidation
path was developed as part of SCADMAN~\cite{scadman}; for \name, we augment it to
recover program-side evidence including routines or function blocks,
controller tags, read/write
dependencies, call relationships, and available state structure. In the
current evaluation, this evidence supports controller-aware studies in
code-rich water and manufacturing cases.

When a validation slice has a suitable temporal specification, \name can
attach an optional Signal Temporal Logic (STL) guardrail evaluated with
RTAMT~\cite{nivckovic2020rtamt}. Given a temporal specification
$\varphi_{\tau}$ and target trace $x$, the guardrail computes robustness
\[
\rho(\varphi_{\tau},x).
\]
The sign and magnitude of this robustness value provide a slice-level
assessment of whether the evaluated trace satisfies the specified temporal
property. In the current evaluation, this mechanism is exercised as bounded
extension evidence for WADI$\rightarrow$WD boundary cases rather than as a
grounding criterion or full-system verification mechanism.

\subsection{Assessment Outputs and Provenance}
\label{sec:impl-outputs}

The final \name assessment package records the grounding decision,
validation-slice metadata, adequacy classification, available consumer
outputs, contextual limitations, and provenance links for the selected SUT.
Depending on the analyses applied to a slice, the package may additionally
contain typed alerts, classification results, provenance records, or an
indication that the available target evidence is insufficient for the
requested evaluation.

We additionally exercise the validation-slice output structure through
prototype downstream adapters. A SCADMAN-motivated classification
adapter~\cite{scadman} applies post-analysis classification to detector
windows, while an ICSTracker-motivated provenance
adapter~\cite{ahmed2025icstracker} expands rule templates over eligible
mappings. These prototypes evaluate additional uses of the \name output
representation and are reported separately in
Appendix~\ref{app:downstream-adapters}.

\name maintains an evidence ledger informed by the provenance representation
used in OTThreat~\cite{paul2023towards}. The ledger records source documents,
extraction runs, canonical threat objects, field-level evidence spans,
provenance tiers, admissibility metadata, consumer-readiness metadata, and
grounding records. It is implemented as a relational SQLite database with
JSON, CSV, and Cypher exports for inspection and analysis. This representation
preserves traceability between source evidence, structured threat abstractions,
and subsequent grounding results.

\section{Evaluation}
\label{sec:eval}

We evaluate \name across a sequence of distinct evidentiary questions.
Starting from heterogeneous source evidence, \name constructs structured threat
abstractions, evaluates them against implemented structural-admissibility
filters on target systems,
materializes accepted groundings as validation slices, and examines subsequent
target-side evaluation where the required execution substrate is available.
The evaluation therefore reports broad continuous-process grounding, a
pinned-snapshot candidate comparison with a bounded assignment audit,
detailed validation-slice studies, downstream consumer compatibility, a
bounded dynamic-realizability study, and extension studies under separate
evidence surfaces and denominators.

The reported measurements correspond to different stages of the methodology.
\emph{Grounding acceptance} indicates that a source threat abstraction
satisfies the deterministic eligibility criteria defined by the target
contract. \emph{Slice adequacy} assesses whether the resulting validation
slice provides the signal coverage, dependency closure, and timing
evaluability required for the intended target-side evaluation.
\emph{Dynamic realizability}, where evaluated, asks whether the grounded
effect can be produced under a specified plant model and command surface.
\emph{Consumer applicability} determines whether a particular downstream
analysis can operate on an eligible slice, while \emph{consumer outcome}
records the result produced by an applicable consumer.

\subsection{Source Semantics and Provenance}
\label{sec:eval-source}

Target-conditioned grounding requires source evidence to be represented as
structured, traceable threat semantics. The evaluation corpus contains 83
canonical source-threat abstractions drawn from 12 source documents spanning
SWaT, WADI, OilTreatment, and
Fischertechnik manufacturing artifacts. Of these, 78 form the
continuous-process grounding corpus: 51 from SWaT, 17 from WADI, and 10 from
OilTreatment. The remaining five Fischertechnik abstractions exercise
manufacturing/program-artifact grounding and are reported separately.

\begin{table*}[t]
\centering
\small
\caption{Source semantics and provenance. \name extracts typed source-threat abstractions and preserves field-level evidence before target grounding. Counts are evaluation artifacts produced by \name, not external benchmark labels.}
\label{tab:source-semantics}
\begingroup
\setlength{\tabcolsep}{4pt}
\renewcommand{\arraystretch}{1.12}
\footnotesize
\begin{tabular}{@{}p{0.26\linewidth}p{0.29\linewidth}p{0.37\linewidth}@{}}
\hline
\textbf{Evaluation object} & \textbf{Result} & \textbf{Interpretation} \\
\hline
Structured threat corpus &
83 threats from 12 documents; 55 high-confidence, 18 medium-confidence, 10 low-confidence. Continuous-process matrix sources: SWaT 51, WADI 17, OilTreatment 10 ($n=78$). Manufacturing extension: Fischertechnik 5. &
Provides typed source semantics across water, oil/chemical, and manufacturing evidence for deterministic target grounding. \\

WADI source deepening &
Naive PDF parsing: 2 rows; \namenospace: 15 A1 threats + 14 A2 attack windows &
Source-specific extraction recovers concrete attack points, effect families, and timing/observability context missed by table scraping. \\

\name evidence ledger &
333 curated threats; 1,947 field-level evidence spans; 100\% core-field coverage &
Our OTThreat-aligned audit layer links paper-facing threat objects to source evidence; it is provenance support, not the grounding algorithm. \\

\hline
\end{tabular}
\endgroup
\end{table*}

Table~\ref{tab:source-semantics} summarizes this source-evidence layer. The
83 abstractions comprise 55 high-, 18 medium-, and 10 low-confidence records.
For WADI, structured extraction refines coarse attack-table entries into
concrete attack points, effect families, attack windows, and observability
context. These abstractions and their associated provenance constitute the
source-side inputs to target-conditioned grounding.
A separate single-reviewer check examined source locations and excerpts for
all 83 paper-facing records.

\name uses OTThreat~\cite{paul2023towards} as part of the lineage informing
its evidence representation while maintaining a separate evidence ledger for
the current framework. The broader ledger contains 333 threat objects and
1,947 field-level evidence spans across 16 source documents. It records source
excerpts, extraction metadata, confidence, and other provenance required to
trace structured abstractions to their supporting evidence. The 333-object
ledger therefore represents a broader evidence and provenance population, not
333 independently validated, grounded, or executable attacks. The evaluation
denominators used below are the 78 continuous-process abstractions and five
manufacturing/program-artifact abstractions, with executable claims restricted
further to the cases actually instantiated and exercised.
Appendix~\ref{app:source-evidence} reports additional source-evidence and
ledger details.

\subsection{Continuous-Process Grounding Coverage}
\label{sec:eval-grounding}

\noindent\textbf{RQ1: Structural grounding.}
We first evaluate the extent to which continuous-process threat abstractions
are structurally groundable across the evaluated target process families. This
analysis considers the 78 continuous-process abstractions derived from SWaT,
WADI, and OilTreatment and applies the deterministic grounding procedure to
water-treatment, water-distribution, hydro/water-energy, and chemical-process
(GRFICS/Tennessee-Eastman, TE) target surfaces. The five Fischertechnik
abstractions are excluded from this matrix because they exercise manufacturing/workcell and program-artifact
semantics and are evaluated separately in
Section~\ref{sec:eval-extensions}.

\begin{figure}[t]
    \centering
    \includegraphics[width=\linewidth]
    {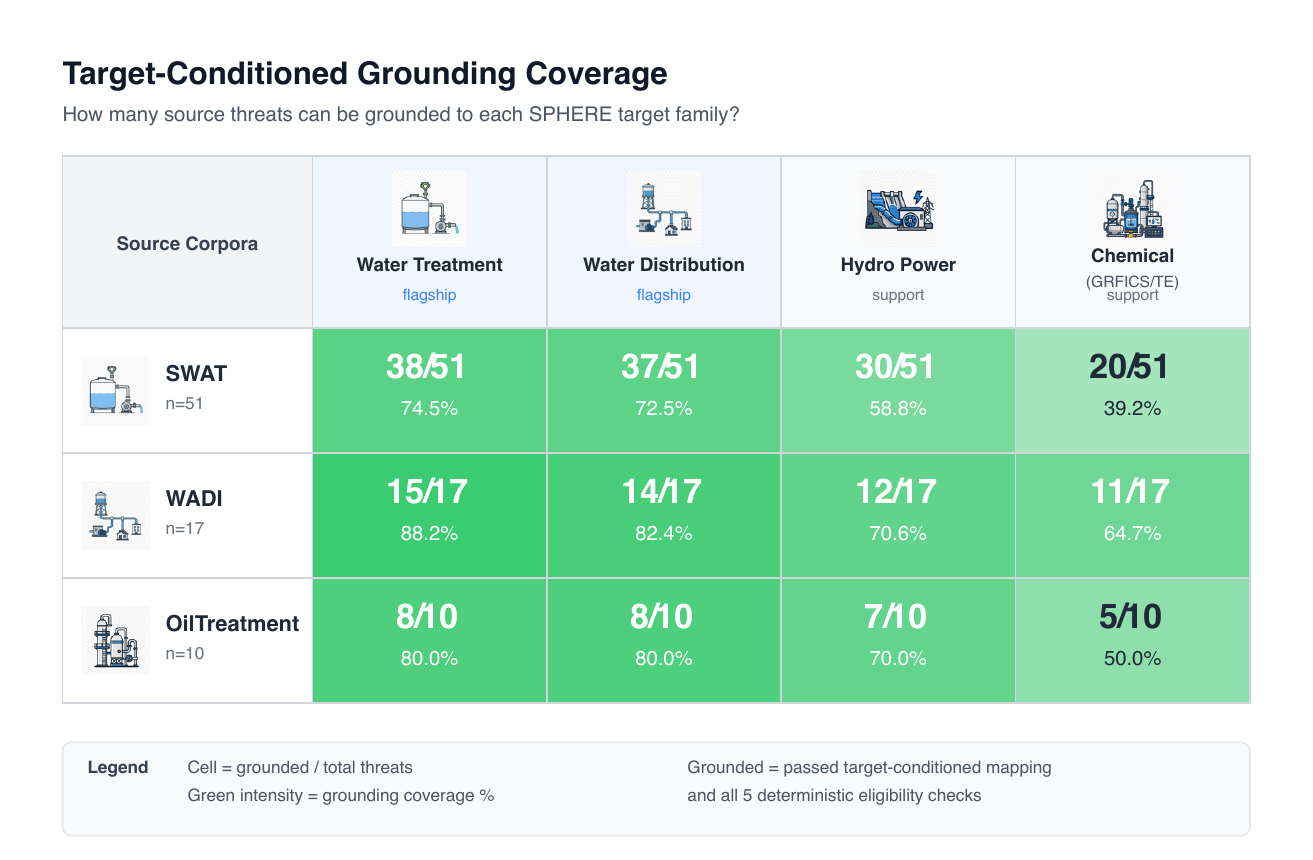}
    \caption{Continuous-process grounding coverage across target SUT families.
    Rows denote source threat corpora and columns denote target process
    families. Each cell reports accepted groundings relative to the
    source-threat denominator for that row. The matrix comprises 78
    continuous-process abstractions from SWaT, WADI, and OilTreatment; the
    five Fischertechnik manufacturing abstractions are evaluated separately
    under a manufacturing/program-artifact denominator.
    Table~\ref{tab:candidate-coverage} reports a separately pinned input
    snapshot.}
    \label{fig:grounding-matrix}
\end{figure}

For each source abstraction $\tau$, the grounder constructs candidate target
mappings $\mu$ and evaluates them using the eligibility predicates defined in
Section~\ref{sec:framework:grounding}: component coverage, implemented
role/type compatibility, source-stage coherence, minimum-tag slice viability,
and rule-surface intersection. Figure~\ref{fig:grounding-matrix} reports the resulting
grounding coverage. Each cell represents the proportion of source
abstractions satisfying these criteria for the corresponding target process
surface. Grounding coverage therefore measures acceptance under the
implemented structural-admissibility filters for the target contract and does
not measure slice adequacy, dynamic realizability, or downstream consumer
performance.

Grounding coverage is highest for water-domain sources evaluated against
water-domain targets and lower for several adjacent process families.
Differences in coverage reflect the eligibility conditions imposed by the
corresponding target contracts, including component coverage, implemented
role/type compatibility, source-stage coherence, minimum-tag slice viability, and intersection
with the declared rule surface. The matrix therefore characterizes the
observed scope and boundaries of structural grounding across the evaluated
source--target combinations.

\noindent\textbf{Pinned-snapshot candidate comparison.}
Because exact input equivalence for Figure~\ref{fig:grounding-matrix} was
not established (Section~\ref{sec:threats-validity}), we ran a separate
comparison on a newly pinned input snapshot that fixes the versions of the
source artifact, grounder, grounding configuration, baselines, runner, four
target contracts, and four target rule files and records their SHA-256
content digests. It
covers the same comparison universe of 78 continuous-process sources and four
targets: 312 source--target pairs and 664 source components, counted once per
target. The snapshot comparison reproduces the per-pair outcomes reported in
Figure~\ref{fig:grounding-matrix} in acceptance, mapped tags, applicable
rules, and all five eligibility dimensions.

On this snapshot, we compare \name's candidate construction with two
deliberately simple baselines. Both receive only source-component
identifiers and roles and target-tag metadata, and each returns at most one
target tag per source component. The literal name baseline requires equality
of case-folded identifiers and tag names with non-alphanumeric characters
removed. The role-only baseline takes the noun preceding a
\texttt{\_sensor} or \texttt{\_actuator} role suffix and requires it as an
exact token in a tag name or description, together with the same tag
direction. Neither baseline uses \name signal aliases, target role
preferences, rule surfaces, or slice checks. All three methods are counted
under the same definition: a pair is \emph{candidate-bearing} when the
method returns at least one assignment.
Table~\ref{tab:candidate-coverage} reports the result. The literal baseline
returned no candidates on any pair; it marks a nomenclature floor rather than
a competing method. The role-only baseline and \name returned candidates for
206 and 222 of the 312 pairs, respectively. They returned identical nonempty
assignments on 44 pairs, and both abstained on 89. These are
candidate-coverage counts, not correctness judgments, and more returned
candidates do not imply better mappings. The 16-pair difference consists of
$+17$ pairs in the Hydro column and $-1$ in WADI$\rightarrow$WT. Because
\name also uses configured signal aliases and role preferences that the
baselines lack, the comparison does not isolate an algorithmic improvement.
Separately, \name accepted 205 of the 312 pairs after its five eligibility
checks; the baselines have no corresponding acceptance stage.
Appendix~\ref{app:mapping-audit} reports the counts for each source-domain
$\times$ target cell.

\begin{table}[t]
\centering
\footnotesize
\caption{Candidate coverage on the separately pinned input snapshot (78 continuous-process sources $\times$ four targets). A pair is
candidate-bearing when the method returns at least one source-component
assignment; components are counted once per target. These are counts of
returned candidates, not correctness or acceptance. \name additionally
accepted 205 of 312 pairs after its five eligibility checks, a separate
measure with no baseline counterpart.}
\label{tab:candidate-coverage}
\begin{tabular}{@{}lrr@{}}
\toprule
\textbf{Candidate method} &
\textbf{Candidate-bearing pairs (/312)} &
\textbf{Mapped source components (/664)} \\
\midrule
Literal name baseline & 0 & 0 \\
Role-only baseline & 206 & 377 \\
\name candidate construction & 222 & 408 \\
\bottomrule
\end{tabular}
\end{table}

\noindent\textbf{Bounded assignment audit.}
Candidate coverage and eligibility are assessed per pair; to examine what
these counts do not show, we audited the individual assignments returned on
24 source--target pairs. These were the two highest-ranked pairs under a
frozen hash order in each source-domain $\times$ target cell, among pairs
with at least one returned assignment. The 24 pairs contain 116 distinct
returned assignments: 61 from the role-only baseline and 65 from \name, of
which 10 were returned by both. The assignments are concentrated: one pair
contributes 42 of them. A human reviewer and two
language-model judgment streams separately labeled each assignment on method-blind evidence cards as
supported, insufficiently evidenced, or unsupported. All three judgment
streams gave the same verdict on only 36 of the 116 assignments, and their
verdict totals differed substantially. These pooled totals are
assignment-level counts clustered within 24 pairs, not 116 independent
cases. The selection is stratified and non-proportional rather than random.
The cards omit raw source documents and some evidence used by \name, which
may contribute to, but is not shown to explain, the disagreement. None of the
streams is an independent oracle. We therefore report the audit only as
diagnostic evidence that candidate coverage should not be read as mapping
correctness. It is not an accuracy estimate or a method ranking.
Appendix~\ref{app:mapping-audit} describes the selection rule, evidence
cards, judgment streams, custody, verdict totals, agreement, and
limitations.

\noindent\textbf{Shared and target-specific inputs.}
Table~\ref{tab:target-config-inventory} lists which inputs the four targets
share and which are configured separately for each one, as recorded in the
pinned snapshot. All targets share the source abstractions, the
candidate-construction and five-check grounding code, the grounding
configuration file, and the comparison runner. Each target supplies its own
tag contract, defense-rule file, and configuration entry for its minimum
slice size, signal aliases, and role-preference tags. This inventory
describes which artifacts must be supplied for a target. We did not measure
onboarding effort or time savings.

\begin{table}[t]
\centering
\footnotesize
\caption{Shared and target-specific inputs in the pinned snapshot.
Entries are taken from the snapshot manifest and the pinned grounding
configuration; they describe artifact structure.}
\label{tab:target-config-inventory}
\begin{tabularx}{\linewidth}{@{}>{\raggedright\arraybackslash}p{0.30\linewidth}>{\raggedright\arraybackslash}X@{}}
\toprule
\textbf{Input} & \textbf{Scope and pinned artifact} \\
\midrule
\multicolumn{2}{@{}l}{\emph{Shared by all four targets (frozen \name evaluation artifacts)}} \\
Source abstractions & One structured source artifact; the same 78
continuous-process records enter every target comparison. \\
Candidate construction and five eligibility checks & One grounder
implementation, unchanged across targets. \\
Grounding configuration & One file with one section per target. \\
Candidate baselines and runner & One baseline module and one comparison runner. \\
\midrule
\multicolumn{2}{@{}l}{\emph{Configured separately for WT, WD, Hydro, and GRFICS/TE}} \\
Target tag contract & Tag metadata (name, type, direction, role, group,
units, description) and declared dependency relations; one contract per
target, from a pinned version of the SPHERE use case. \\
Defense-rule surface & One rule file per target, from a pinned version of the
CPS enclave model. \\
Per-target configuration entries & Contract and rule references, minimum
slice size, signal aliases, and role-preference tags; populated for all four
targets. \\
\bottomrule
\end{tabularx}
\end{table}

\subsection{Testing Grounded Water-Domain Cases}
\label{sec:eval-slices}

\noindent\textbf{RQ2: Slice evidence.}
Grounding acceptance does not by itself establish that the resulting
target-side representation provides sufficient execution evidence for
downstream evaluation. We therefore examine two detailed water-domain studies,
SWaT$\rightarrow$WT and WADI$\rightarrow$WD, by materializing selected
accepted groundings as executable validation slices. The corresponding study
bundles provide target signals, dependency relationships, nominal traces,
controlled perturbation cases, timing information, and the metadata required
to assess slice adequacy and downstream outcomes.

\noindent\textbf{Case selection and execution.}
The five detailed cases are not a random sample of the 78 continuous-process
source abstractions. Of these abstractions, 61 ground to WT and 59 to WD
(Figure~\ref{fig:grounding-matrix}). We construct an executable validation case
only when the mapped manipulated and consequence signals and the required
dependencies are represented in the target study bundle and when both nominal
evidence and a viable controlled perturbation surface are available. From this
eligible set, we report five cases selected to span multiple effect families
and to include a boundary case in addition to clearer downstream responses
(Table~\ref{tab:flagship-slices}). These experiments therefore examine a
selected subset for which the required target-side execution substrate was
available.
All five reported slices satisfy the declared adequacy requirements for signal
coverage, dependency closure, and timing evaluability.

\begin{table*}[t]
\centering
\small
\caption{Water-domain validation-slice outcomes. Controlled target-side
perturbations probe slices supported by accepted groundings. Outcomes are labeled as clean,
nominal-confounded, or near-threshold.}
\label{tab:flagship-slices}
\begingroup
\setlength{\tabcolsep}{4pt}
\renewcommand{\arraystretch}{1.12}
\footnotesize
\begin{tabular}{@{}p{0.34\linewidth}p{0.58\linewidth}@{}}
\hline
\textbf{Grounding and target case} & \textbf{Target-side result and interpretation} \\
\hline

\textbf{SWaT$\rightarrow$WT: Tank-level high spoof.}
Positive level offset on the WT tank slice; a target-side probe, not a
direction-matched replay of SWaT attack 36. &
Nominal-subtracted analysis yields three invariant violations: two rate-of-change and one mass-balance. This is the cleanest WT invariant case. \\

\textbf{SWaT$\rightarrow$WT: Pump-flow low spoof.}
Pump/flow suppression instantiated on the WT slice. &
Raw analysis yields nine correlation violations, but nominal-subtracted analysis yields zero. We treat this as nominal-confounded and do not count it as a clean invariant success. \\

\textbf{WADI$\rightarrow$WD: Supply-flow low spoof.}
Low-flow effect instantiated on the WD supply slice. &
No baseline invariant violations occur; the first firing appears only at the $-64$ perturbation boundary with shallow range/correlation evidence. We report this as a near-threshold boundary case. \\

\textbf{WADI$\rightarrow$WD: Supply-flow high idle.}
High-flow/pressure effect instantiated during an idle supply condition. &
The slice produces 11 range violations, giving a clean WD flow/pressure case. \\

\textbf{WADI$\rightarrow$WD: NaOCl level step.}
Dosing/storage-level perturbation instantiated on the WD chemical slice. &
The slice produces three mass-balance/rate-of-change violations. This is a clean WD dosing case, although it is outside the GeCo predictive surface. \\

\hline
\end{tabular}
\endgroup
\end{table*}

\noindent\textbf{Source grounding versus WT probe.}
The SWaT attack 36 record in Figure~\ref{fig:semantic-example} lists only
\texttt{LIT-101} as manipulated and reports tank overflow. Its v1 structural
grounding maps that role to the WT signal \texttt{RW\_Tank\_Level}; the WT
slice includes
\texttt{RW\_Tank\_PR\_Valve} as target context, not as a mapping of
\texttt{MV-101} from this source record. The type, stage, structure, and
rule-surface eligibility checks are satisfied.

The separate WT high-spoof test applies a positive offset to
\texttt{RW\_Tank\_Level}. Attack 36 instead sets \texttt{LIT-101} below the low-low (LL) alarm level,
so this test probes the mapped WT signal but does not replay the source
attack's direction. The native invariant consumer yields three
nominal-subtracted violations: two rate-of-change and one mass-balance. This
demonstrates a response to the controlled WT perturbation, not adversary
access, exploit feasibility, or equivalent behavior under other plant
dynamics.

The SWaT pump-flow case further illustrates the separation between grounding,
adequacy, and consumer outcome. The slice satisfies both grounding and
adequacy criteria, but all nine raw correlation violations also occur in the
nominal trace and disappear after nominal subtraction. The consumer outcome
is therefore classified as \emph{nominal-confounded}, while the grounding and
slice-adequacy classifications remain unchanged. The WADI$\rightarrow$WD cases
illustrate the same distinction: an adequately grounded slice may produce a
response that survives the nominal baseline or may remain near the selected
consumer's decision boundary without changing the underlying grounding.

These studies characterize validation slices as an intermediate
target-conditioned representation between source-derived threat semantics and
downstream evaluation. The slices preserve the signals, dependencies, and
context required for the corresponding analyses while maintaining separate
grounding, adequacy, and consumer-outcome classifications. The perturbations
are controlled modifications of simulator-role signals applied to otherwise
nominal SPHERE executions. Appendix~\ref{app:water-slice-details} reports the per-case adequacy
metadata and raw and nominal-subtracted outcomes, while
Section~\ref{sec:threats-validity} discusses the resulting evidentiary
limitations.

\subsection{Consumer Compatibility over Shared Slices}
\label{sec:eval-consumers}

\noindent\textbf{RQ3: Interoperability.}
We next evaluate whether a common validation-slice representation can provide
the information required by multiple downstream analysis styles.
Table~\ref{tab:consumer-utility} summarizes four bounded consumer lanes
evaluated over the WT/WD slices: the native invariant-family consumer, a
bounded GeCo-style predictive lane~\cite{wolsing2025geco}, a bounded
SAIN-style state-aware lane~\cite{abbas2024sain}, and a bounded
SCAPHY-style phase-aware lane~\cite{ike2022scaphy}. The latter three are
\name-native implementations of selected analysis patterns motivated by the
corresponding systems; they do not execute the published implementations.

\begin{table}[t]
\centering
\footnotesize
\caption{Bounded consumer results over shared validation slices. Each row applies a different bounded analysis style to the same WT/WD slice artifacts.}
\label{tab:consumer-utility}
\begingroup
\setlength{\tabcolsep}{4pt}
\renewcommand{\arraystretch}{1.12}
\footnotesize
\begin{tabular}{@{}>{\raggedright\arraybackslash}p{0.20\linewidth}>{\raggedright\arraybackslash}p{0.25\linewidth}>{\raggedright\arraybackslash}p{0.25\linewidth}>{\raggedright\arraybackslash}p{0.22\linewidth}@{}}
\hline
\textbf{Consumer family} & \textbf{WT slice results} & \textbf{WD slice results} & \textbf{Role in evaluation} \\
\hline

Invariant checks &
Raw 2/2; 1 clean case remains after nominal subtraction. &
2/3; high-flow and NaOCl are strong, low-flow is partial. &
Native slice consumer; preserves WT nominal caveat. \\

GeCo-style predictive~\cite{wolsing2025geco} &
2/2 hits; nominal FPR 0.0; mean F1 0.4167. &
2/3 hits; nominal FPR 0.0; mean F1 0.1025; NaOCl outside modeled surface. &
Bounded predictive compatibility lane. \\

SAIN-style state-aware~\cite{abbas2024sain} &
2/2 hits; nominal FPR improves from 0.0167 to 0.0; mean F1 0.66. &
3/3 hits; nominal FPR 0.0; mean F1 1.0. &
Bounded state-aware compatibility lane. \\

SCAPHY-style phase-aware~\cite{ike2022scaphy} &
2/2 hits; nominal FPR 0.0; mean F1 0.5661. &
3/3 hits; recovers low-flow case; nominal FPR 0.0; mean F1 0.57. &
Bounded phase/context-aware compatibility lane. \\
\hline
\end{tabular}
\endgroup
\end{table}

The invariant-family lane provides the native rule-based evaluation of the
WT/WD validation slices, including the nominal-confounded WT pump-flow case.
The bounded GeCo-style lane identifies both WT cases and two of the three WD
cases at a target nominal false-positive rate (FPR) of 0.0; the NaOCl dosing
case lies outside its modeled surface. The bounded SAIN-style lane removes the nominal
false-positive behavior observed for WT while retaining the reported WT and WD
coverage. The bounded SCAPHY-style lane additionally identifies the WD
low-flow boundary case, increasing the observed WD coverage from 2/3 to 3/3
within that bounded implementation.

Across these lanes, the validation-slice representation supplies signals and
context used by several downstream analysis styles, including invariant
evaluation, predictive residual analysis, state-conditioned thresholds, and
phase-conditioned scoring. These results provide evidence of consumer
compatibility with the shared slice representation under the evaluated
implementations. Implementation scope and consumer-specific details are reported in
Appendix~\ref{app:consumer-implementations}.

\subsection{Bounded-Support Validation Slices: Hydro and Chemical-Process Targets}
\label{sec:eval-hydro-grfics}

We next examine nine bounded-support cases spanning
SWaT$\rightarrow$Hydro, WADI$\rightarrow$GRFICS, and
EPIC$\rightarrow$Hydro. These cases use the same target-contract and
controlled-perturbation methodology as the primary validation studies
(Section~\ref{sec:impl-target}). Because corresponding nominal-subtracted
baselines are unavailable, the resulting evidence is limited to
validation-slice executability and observed bounded-consumer response rather
than the stronger outcome interpretation available for the WT/WD flagship
cases. Appendix~\ref{app:hydro-grfics-details} reports the full adequacy and
timing metadata.

Across the nine cases, pressure- and flow-related perturbations produce less
consistent activation of dependency-based rules than of range rules. A
similar pattern is observed for the WD low-flow case
(Section~\ref{sec:eval-slices}). Interpretation of these cases is constrained
because several Hydro cases share a target execution bundle.
Consequently, variation in observed downstream rule response does not modify
the structural grounding or slice-adequacy classifications. These cases
provide evidence of validation-slice executability and bounded consumer
compatibility across the evaluated target families without establishing
target-independent behavior or validation against independently executed
attacks.

\begin{table}[t]
\centering
\footnotesize
\caption{Bounded-support validation slices on hydro and chemical-process
targets. All nine satisfy the reported slice-adequacy requirements but lack
nominal-subtracted baselines; the observed responses therefore characterize
slice executability and bounded consumer compatibility rather than clean
consumer outcomes. $^\dagger$Shared Hydro executions provide convergent
source-side grounding evidence while sharing the same target-side execution
evidence.}
\label{tab:hydro-grfics-slices}
\begin{tabularx}{\linewidth}
{@{}p{0.16\linewidth}p{0.19\linewidth}p{0.19\linewidth}X@{}}
\toprule
\textbf{Case} & \textbf{Effect / perturbation} &
\textbf{Observed response} & \textbf{Interpretation} \\
\midrule

SWaT$\to$Hydro: reservoir level &
overflow, \texttt{HY\_Res\_Level} $+20$ &
range, rate-of-change fire (2/2) &
Both declared rule types respond \\

SWaT$\to$Hydro: turbine speed &
speed excursion, \texttt{HY\_Speed\_Pct} $+70$ &
range, rate-of-change fire (2/2) &
Both declared rule types respond \\

SWaT$\to$Hydro: flow &
pressure/flow, \texttt{HY\_Flow} $-90$ &
range fires only (1/3) &
Limited observed dependency-rule response \\

WADI$\to$GRFICS: tank pressure &
pressure/flow, \texttt{TE\_Tank\_Pressure} $+1500$ &
rate-of-change fires only (1/2) &
Limited observed rule response \\

WADI$\to$GRFICS: tank level &
overflow, \texttt{TE\_Tank\_Level} $+40$ &
rate-of-change and causality responses &
Mixed observed rule response \\

WADI$\to$GRFICS: feed-1 flow &
pressure/flow, \texttt{TE\_Feed1\_Flow} $-250$ &
causality, range fire (2/3) &
Broadest observed GRFICS rule response \\

EPIC$\to$Hydro: turbine speed &
speed excursion, \texttt{HY\_Speed\_Pct} $+70$ &
range, rate-of-change fire (2/2), identical to SWaT-sourced case &
Shared target execution$^\dagger$ \\

EPIC$\to$Hydro: generated power &
speed excursion, \texttt{HY\_Power\_MW} $-40$ &
range, correlation fire (2/2) &
Both declared rule types respond \\

EPIC$\to$Hydro: flow &
pressure/flow, \texttt{HY\_Flow} $-90$ &
range fires only (1/3), identical to SWaT-sourced case &
Limited observed dependency-rule response; shared execution$^\dagger$ \\

\bottomrule
\end{tabularx}
\end{table}

\subsection{Upstream GeCo as a Consumer}
\label{sec:eval-geco-real}

The GeCo-style lane in Section~\ref{sec:eval-consumers} is a bounded
\name-native implementation of a predictive analysis pattern motivated by
GeCo. We separately evaluate compatibility with the published upstream GeCo
implementation by executing the unmodified artifact, including its released
per-case retraining and Industrial Protocol Abstraction Layer (IPAL) evaluation
pipeline, over nine
\name-grounded slices spanning three studies. For each case, the evaluation
uses 60 target rows, 21 of which correspond to the perturbation window.
The SPHERE-derived bundles are represented in IPAL state format, after which
the upstream GeCo pipeline performs its released training and scoring
procedure without modification.

The upstream evaluation additionally reports eTaF1 (enhanced time-series-aware F1), the time-aware metric
used by GeCo. As an artifact-level consistency check, we also executed the
released evaluation on GeCo's native benchmark material. The WADI result
matches the retained comparison baseline available with the evaluated
artifact.

The nine slice executions do not correspond to nine independent target-side
observations. Two cases share the same Hydro target execution and nominal
evidence as corresponding cases derived from different source-side
provenance, and consequently produce identical GeCo measurements. These cases
therefore provide convergent grounding evidence from distinct source-side
provenance while sharing target-side execution evidence.

Across the nine executions, GeCo obtains recall $1.0$ with
$\mathrm{tp}=21$ and $\mathrm{fn}=0$ for each case, while false positives
range from 2 to 3. Only two distinct eTaF1 values occur across the nine cases.
These measurements therefore provide little case-level differentiation among
the finer interpretations reported in
Table~\ref{tab:hydro-grfics-slices}. The concentration of observed outcomes
is consistent with the large controlled perturbations and small per-case
evaluation setting used here. The controlled-perturbation limitations
described in Section~\ref{sec:threats-validity} continue to apply.

This experiment establishes that the evaluated \name validation slices can
serve as inputs to the unmodified upstream GeCo pipeline and can be scored
using its released time-aware evaluation procedure. The evidence therefore
supports downstream consumer compatibility for the evaluated cases. It does
not establish transfer quality, case-level discrimination, independent
target-side replication, comparative superiority, or validation against
independently executed attacks.

\subsection{Dynamic Realizability: A Bounded Example}
\label{sec:eval-realizability}

\noindent\textbf{RQ4: Dynamic realizability.}
Structural grounding does not establish that a grounded effect can be produced
through a particular plant model and legitimate command surface. We examine
this distinction for three frozen \name groundings on the
GRFICS/Tennessee-Eastman target: one pressure-family objective
(\texttt{pressure\_flow\_perturbation}, $P>3100$) and two level-family
objectives (\texttt{underflow\_depletion}, $L\le5$;
\texttt{overflow\_accumulation}, $L\ge95$). For these cases, we use a
paper-derived reproduction of the physics-guided search method described by
ICSFlux~\cite{11573458} as a separate check of dynamic realizability.

Each search objective is compiled from a frozen grounded threat and specifies
the corresponding violation condition over the evaluated observable and
command surface. The search procedure is separate from the controlled
additive perturbation used to exercise the validation slices in the preceding
studies. This experiment therefore asks a different question: whether an
already-grounded objective can be reached under a specified model and command
surface.

Against the plant model and command surface specified in the ICSFlux appendix,
the pressure objective is reached, whereas the two level objectives are not
reached under the evaluated search and extremal-command campaign. In the
additional extremal campaign, every level-relevant command is held at its
corresponding extreme for 400 simulated hours while the controller retains
the level within the non-violating band. Under the live OpenPLC-driven plant
configuration used for this bounded study, all three objectives are reached.
These observations are limited to three objectives on a single target and
provide evidence that structural grounding and model-specific dynamic
realizability are distinct properties. The empirical negative result is
bounded by the evaluated model, command alphabet, and campaign. The difference between the two execution
settings should not be interpreted as a property of the grounding decision or
of the target independently of those conditions.

\subsection{Extension Evidence and Boundaries}
\label{sec:eval-extensions}

Finally, we examine whether the \name evidence representation can support
additional artifact types and analysis surfaces beyond the primary
continuous-process studies. These experiments exercise different denominators
and evidence forms and are therefore reported separately rather than combined
with the continuous-process grounding percentages or the primary
validation-slice results.

CISS~\cite{itrust2019ciss} contributes additional water-treatment attack
semantics. The Fischertechnik TXT$\rightarrow$Siemens S7 study exercises
manufacturing/program-artifact grounding using program, tag, and state
evidence. CrossPLC exercises optional controller-aware enrichment by exposing
controller-side tags, routines or function blocks, read/write dependencies,
call relations, and available state structure in code-rich water and
manufacturing cases. These results do not alter the denominator of the
continuous-process grounding matrix. STL/RTAMT
guardrails~\cite{nivckovic2020rtamt} provide bounded temporal robustness
evaluation for WADI$\rightarrow$WD boundary cases. EPIC/Hydro~\cite{epic} and
PowerDuck Generic Object Oriented Substation Event
(GOOSE) data~\cite{powerduck} contribute power- and protocol-oriented
evidence, including GOOSE flooding, insertion, replay, and suppression traces,
while VetPLC-style artifacts~\cite{vetplc} contribute event and timing
obligations for manufacturing-oriented settings.

Taken together, these studies show that the \name representation can be
instantiated over additional evidence types and mechanisms. They do not
provide the same target-side evidence as the primary WT/WD studies and are
therefore interpreted under their own experimental scope.
Appendix~\ref{app:extension-evidence} reports the corresponding
extension-evidence ledger.

\section{Discussion}
\label{sec:discussion}

\noindent\textbf{Grounding and downstream evaluation.}
The grounding matrix answers a narrow question: which source abstractions pass
the implemented component-coverage, role/type, source-stage, minimum-tag, and
rule-intersection filters for each target contract? A rejected candidate fails one or more grounding
checks; a partial or insufficient slice instead identifies a limitation in
the evidence available after grounding.

The validation slice records the mapped paths, signals, dependencies, timing
assumptions, and context associated with an accepted result. Subsequent
analyses address different questions about that record. For example, a strong slice may still
produce a \emph{nominal-confounded} consumer outcome because the same violation
appears in the nominal trace, or it may not expose the inputs expected by a
particular consumer.

\noindent\textbf{Complementarity with downstream analyses.}
\name produces target-conditioned validation artifacts rather than a new
detector or provenance-analysis system. The bounded GeCo-, SAIN-, and SCAPHY-style
lanes (Section~\ref{sec:impl-consumers}) are \name-native implementations of
selected analysis patterns motivated by the corresponding systems and are used
to evaluate whether the validation-slice representation provides the required
consumer inputs. The upstream GeCo study
(Section~\ref{sec:eval-geco-real}) separately demonstrates that the evaluated
\name-grounded slices can be processed by the released GeCo pipeline. Its
evidentiary scope is consumer compatibility rather than transfer quality,
case-level discrimination, or comparative detector performance.

The classification- and provenance-oriented adapters in
Appendix~\ref{app:downstream-adapters} exercise additional output interfaces
motivated by SCADMAN and ICSTracker without executing those published systems.
Together, these studies exercise several output interfaces; only the GeCo
study establishes compatibility with a released upstream implementation.

\noindent\textbf{Evidence quality and grounding boundaries.}
The quality of \name's output depends on the completeness and accuracy of the
available source and target evidence. Missing process diagrams, undocumented
bypasses, stale tag names, unit or type mismatches, and vendor-specific aliases
may prevent otherwise plausible candidate mappings from satisfying the
grounding criteria or may reduce the adequacy of the resulting slice. \name
therefore retains field-level provenance and explicit eligibility outcomes
rather than filling missing evidence through unconstrained inference. This
design makes the evidentiary basis and residual uncertainty of each result
inspectable, which is particularly important for legacy systems with
incomplete or inconsistent documentation.

\noindent\textbf{Optional enrichment paths.}
CrossPLC and STL/RTAMT provide optional evidence for particular studies without
changing the requirements of the core grounding procedure. In the
water-treatment controller study, CrossPLC contributes controller-side
evidence for role and dependency recovery while leaving the
continuous-process grounding denominator unchanged. The STL/RTAMT study
evaluates temporal robustness around the WADI$\rightarrow$WD boundary case
and therefore supplements slice-level evaluation rather than serving as an
alternative grounding mechanism. These studies illustrate how additional
controller-code and temporal-logic evidence was used in the evaluated cases
without making either source a prerequisite for grounding.

\noindent\textbf{Scope of extension evidence.}
The extension studies in Section~\ref{sec:eval-extensions} exercise evidence
surfaces and denominators distinct from the continuous-process grounding
matrix. These studies include manufacturing and program-artifact grounding,
controller-code enrichment, protocol traces, and temporal obligations.
Accordingly, their results are reported separately and are not included in the
continuous-process grounding percentages or interpreted as equivalent to the
primary WT/WD executable studies.

\noindent\textbf{Security and deployment considerations.}
Grounding over physical roles, process context, dependencies, observable
signals, and declared rule surfaces reduces dependence on exact source--target
tag-name correspondence. The procedure nevertheless remains sensitive to the
integrity and completeness of its input evidence. Incorrect or outdated
manuals, misleading engineering diagrams, stale configuration information,
or adversarially manipulated evidence can affect source abstraction or target
contract construction. Provenance tracking, schema validation, and
cross-source corroboration provide partial safeguards, while stronger
mechanisms for evidence integrity and attestation remain deployment concerns.

Practical deployment also requires engineering effort beyond the grounding
algorithm itself, including alias reconciliation, unit and type normalization,
timestamp alignment, target-contract construction, and maintenance of
target-side dependency information. These activities affect the coverage and
adequacy of the resulting validation slices and should therefore be treated as
part of the deployment boundary.

\noindent\textbf{Scope of deterministic grounding.}
The deterministic claim of \name applies only after the source abstraction,
vocabulary and schema version, and target contract have been fixed. Source
evidence selection, semantic abstraction, confidence assignment, and
target-contract construction may involve analyst interpretation or
source-specific parsing and are outside this deterministic claim. In
addition, the current grounding vocabulary is narrower than the extraction
vocabulary (Table~\ref{tab:semantic-constructs}), so some valid source
abstractions fall outside the effect-family scope currently supported by the
target contracts.

\noindent\textbf{Interpretation of controlled-perturbation evidence.}
The five flagship SWaT$\rightarrow$WT/WADI$\rightarrow$WD cases and the nine
bounded-support Hydro/GRFICS cases
(Section~\ref{sec:eval-hydro-grfics}) use controlled additive perturbations
applied to simulator-role signals during otherwise nominal executions. These
perturbations provide evidence that the corresponding validation slices can be
executed under the evaluated target configurations and that the relevant
target-side signals and dependencies exhibit observable responses.

\noindent\textbf{Structural grounding and dynamic realizability.}
Structural grounding establishes that a source threat abstraction satisfies
the implemented component-coverage, role/type, source-stage, minimum-tag, and
rule-intersection filters. Producing the corresponding effect through a
particular plant model and command surface is a separate question.

The bounded study in Section~\ref{sec:eval-realizability} illustrates this
distinction for three frozen groundings. Their realizability differs between
the analytical plant model used by the paper-derived search reproduction and
the evaluated OpenPLC-driven execution configuration. This result provides a
bounded example in which structural grounding remains unchanged while
model-specific realizability differs.

\subsection{Threats to Validity}
\label{sec:threats-validity}

\noindent\textbf{Construct validity.}
The central construct evaluated in this work is target-conditioned grounding,
defined as acceptance under five eligibility predicates against a fixed source
abstraction, vocabulary and schema version, and target contract. The validity
of this construct therefore depends on both the representational vocabulary
and the selected grounding criteria. The effect-family, role, and rule-surface
representation was developed from the source and target corpora considered in
this study and may not capture threat semantics outside those domains. The
reported grounding evaluation covers only five of the seven effect families
available during source abstraction; abstractions assigned to the two
remaining families fall outside the reported grounding surface.

The five eligibility predicates
($\chi_{\text{map}}, \chi_{\text{type}}, \chi_{\text{stage}},
\chi_{\text{slice}}, \chi_{\text{rule}}$) operationalize structural
admissibility for the evaluated target contracts but are not derived from an
external standard. Alternative vocabularies, mapping procedures, or
eligibility criteria could therefore produce different grounding decisions.
The implemented role/type check is coarse rather than general physical-unit
reasoning. Component coverage is existence-based, stage coherence uses
source-declared stage labels, slice viability uses a minimum expanded-tag
count, and rule relevance uses tag intersection. The reported grounding
matrix therefore measures acceptance under these implemented filters, not
independently adjudicated mapping correctness. Finally, grounding and
validation-slice execution do not establish that an equivalent exploit would
succeed against deployed target control logic.

The candidate comparison in Section~\ref{sec:eval-grounding} uses a
separately pinned input snapshot. The snapshot comparison reproduces the
reported per-pair outcomes in Figure~\ref{fig:grounding-matrix}, although
exact input equivalence was not established.

Candidate coverage, five-check acceptance, and mapping correctness are
distinct constructs; the first two are assessed per pair, while the audit
judges individual assignments. The two comparison baselines are deliberately
simple, name- and role-based procedures that establish context; the
evaluation does not compare \name against stronger mapping approaches.
Complementary
benchmarks such as PLCBench~\cite{zhou2026plcbench} address a different
question and were not run head-to-head.

\noindent\textbf{Internal validity.}
Source-abstraction construction involves analyst judgment. The evaluation
does not include independent semantic adjudication of the source abstractions
or a multi-analyst study. Consequently, it does not provide an inter-annotator
agreement measure or establish that independent analysts would construct
identical abstractions from the same source evidence. The earlier review of
the 83 paper-facing source records checked source locations and excerpts
only, not source--target mapping correctness.

The bounded assignment audit (Appendix~\ref{app:mapping-audit}) judges
returned source--target assignments, not source abstractions. It has its own
limitations. Its 24 pairs were selected by a rule that excludes
both-abstain pairs and takes two pairs per cell regardless of cell size, so
the selection is stratified and non-proportional; the protocol's intended
method-agreement/disagreement enrichment did not take effect because the
literal baseline returned no assignments. Its 116 assignments are clustered
within the 24 pairs: one pair contributes 42 and one cell contributes 47.
Pooled verdict counts therefore weight pairs unequally and are not counts of
independent cases. The two model streams come from
the same provider. The instructions supplied the three verdict labels and
required a brief reason, but no finer written decision rules or operational
definitions were found. The model-task wording was recovered retrospectively;
no separately preserved plaintext prompt was identified, and temperature,
seed, and service tier were not recorded. Evidence cards omit raw source documents
and some evidence available to \name; this may contribute to the observed
disagreement but is not established as its cause. The human-review stream's blinding relied on
procedural self-custody rather than an independently enforced separation.
The audit therefore supports only descriptive
caution about candidate mappings, not accuracy estimates or method rankings.

The detailed validation studies are also subject to selection effects. The five
flagship cases and nine bounded-support cases were selected from cases for
which the required target contract, observable signals, dependency structure,
and viable perturbation surface were available.
Effect-family diversity and the inclusion of a boundary case also influenced
case selection.

The target-side perturbations are controlled additive offsets. Large offsets may be easier for
threshold- or residual-based consumers to identify than subtler, gradual, or
dynamically generated manipulations. This limitation also affects the
interpretation of the upstream GeCo results
(Section~\ref{sec:eval-geco-real}), which provide limited differentiation
across the evaluated cases. The separate realizability study reduces reliance
on the same additive perturbation mechanism for three selected objectives.

\noindent\textbf{External validity.}
The most detailed executable evidence remains concentrated in water-domain
groundings: SWaT into a water-treatment target and WADI into a
water-distribution target within the SPHERE experimental environment. The
bounded-support Hydro and GRFICS chemical-process cases in
Section~\ref{sec:eval-hydro-grfics} extend validation-slice construction and
controlled execution to additional process families, but they remain
research-environment evaluations.

The evaluation therefore does not establish generalization to ICS families
outside the evaluated water, energy, chemical-process, and bounded
manufacturing settings; to substantially different control architectures; or
to targets for which the required role, signal, dependency, or nominal
evidence cannot be constructed. Incomplete target evidence is reflected in
failed grounding criteria or partial/insufficient slice adequacy rather than
treated as evidence of transferability.

\section{Related Work}
\label{sec:related}

\noindent\textbf{Model-Based Cyber-Physical System (CPS) Safety and Security
Coengineering.}
System Aware and Mission Aware cybersecurity use systems-engineering artifacts
to connect mission objectives, system structure, attack vectors, unsafe
actions, hazards, resilience modes, and sentinel-based
monitoring~\cite{jones2011systemaware,jones2012systemaware,
carter2018missioncentric,carter2019preliminary,fleming2021resiliency}.
Bakirtzis et al. extend this perspective through an ontological metamodel that
relates model-based systems-engineering entities, including components, links,
functions, control actions, feedback, and operational context, to
safety-, security-, and resilience-oriented concepts such as attack vectors,
loss scenarios, hazards, unsafe actions, resilient modes, and
sentinels~\cite{bakirtzis2022metamodel}. Related approaches further connect
system models with attack-vector or vulnerability knowledge to support
design-time security analysis~\cite{bakirtzis2018modelbased,bakirtzis2020cybok}.

This body of work emphasizes design-time traceability among architecture,
functions, mission objectives, attack vectors, losses, and resilience
decisions. \name addresses a complementary problem: target-conditioned
grounding of threat evidence documented for one ICS onto a selected target
SUT. Rather than requiring a complete mission-engineering or system model,
\name evaluates whether a structured source threat abstraction satisfies the
target contract and, when it does, constructs a validation slice containing
the manipulated and consequence paths, observability requirements, and
target-side evaluation context required for subsequent analysis.

\noindent\textbf{ICS Threat Intelligence, Ontologies, and Evidence
Representations.}
Security ontologies and knowledge graphs provide structured representations of
cyber--physical threat intelligence. Prior work includes industrial Internet
of Things (IoT) ontologies~\cite{mozzaquatro2016ontology}, cyber threat
intelligence (CTI) knowledge-graph construction using distant
supervision~\cite{shen2020data}, LLM-assisted CTI extraction
systems~\cite{cheng_ctinexus_2025}, and representations oriented to
operational technology (OT), such as OTThreat~\cite{paul2023towards}. These
approaches provide vocabularies, extraction mechanisms, and structured representations for organizing threat,
asset, and evidence information.

\name builds on this representation-oriented foundation by connecting
structured source threat abstractions to explicit target contracts. Its
evidence ledger preserves provenance across source evidence, abstraction
fields, candidate mappings, grounding decisions, and validation-slice
construction. Grounding further requires candidate mappings to satisfy the
implemented component-coverage, role/type, source-stage, minimum-tag, and
rule-intersection criteria. The distinction is therefore between
representing source-side threat semantics and evaluating their structural
admissibility on a particular target.

\noindent\textbf{Target-Side Detection, Monitoring, and Provenance Consumers.}
A broad class of ICS security systems performs analysis after the relevant
target signals, controller behavior, or execution context have been
characterized. GeCo learns discrete state-space models from benign traces and
detects departures from learned behavior~\cite{wolsing2025geco}. SAIN derives
state-aware invariants from PLC program states and transitions~\cite{abbas2024sain}.
SCAPHY relates execution phases on supervisory control and data acquisition
(SCADA) hosts to physical behavior~\cite{ike2022scaphy}. SCADMAN evaluates
control-loop consistency across distributed controllers~\cite{scadman}, while ICSTracker reconstructs
provenance-oriented causal relationships for ICS
forensics~\cite{ahmed2025icstracker}.

These systems focus on downstream analysis over an instrumented or otherwise
characterized target. \name instead addresses the preceding problem of
constructing target-conditioned context from externally documented threat
semantics. Its validation slices identify manipulated and consequence paths,
observable signals, dependencies, and consumer-relevant context that can then
be exposed to downstream analyses.

Sections~\ref{sec:eval-consumers} and \ref{sec:eval-geco-real} examine this
relationship through bounded \name-native consumer lanes and the released
upstream GeCo implementation, respectively.
Appendix~\ref{app:downstream-adapters} additionally evaluates prototype output
interfaces motivated by SCADMAN and ICSTracker. These studies have different
evidentiary scopes and are used to assess downstream compatibility rather than
comparative detector performance or fidelity across systems.

\noindent\textbf{PLC Analysis, Code-Level Testing, and Formal Guardrails.}
PLC-oriented analysis and testing techniques provide assurance at the
controller-program and scan-cycle levels. Attack-pattern mining, static or
dynamic program analysis, and PLC fuzzing can expose implementation-level
weaknesses or execution behaviors~\cite{umer_attack_2025,villa_ics-quartz_2025}.
ICSQuartz, for example, provides a vendor-agnostic analysis path for
IEC~61131-3 Structured Text through a RuSTy-to-LLVM
toolchain~\cite{villa_ics-quartz_2025}.

PLCBench evaluates whether autonomous agents can turn PLC interaction into
sustained physical impact in real-PLC hardware-in-the-loop
settings~\cite{zhou2026plcbench}. \name instead evaluates whether externally
documented threat semantics are structurally admissible on a selected target;
the two questions are complementary.

CrossPLC serves a different role within \name. When controller artifacts are
available, it lifts heterogeneous PLC project formats into a common
representation of tags, routines or function blocks, read/write
relationships, call relations, and available state information. This
representation provides optional controller-level evidence for role recovery,
dependency recovery, and validation-slice construction, but is not required by
the core target-grounding procedure.

Temporal-logic monitoring provides another complementary analysis surface.
RTAMT supports robustness-based evaluation of cyber--physical temporal
specifications~\cite{nivckovic2020rtamt}. \name uses STL/RTAMT as an optional
slice-level mechanism when a grounded validation slice exposes the signals and
thresholds required by a temporal specification. These checks supplement
target-side evaluation and remain separate from grounding acceptance.

\noindent\textbf{Physics-Guided Target Search.}
ICSFlux~\cite{11573458} searches for legitimate command sequences that drive a
target toward a specified violation condition given a black-box controller, a
physical model, and an explicit safety/security constraint. \name addresses a
different stage: it determines whether externally documented threat semantics
are structurally admissible on a selected target. In relation to the three
ICSFlux inputs, \name provides information relevant to formulating the
violation constraint. The two approaches therefore address complementary questions of
target-conditioned threat grounding and model-specific dynamic realizability.

Section~\ref{sec:eval-realizability} examines this distinction through a
bounded study that applies a paper-derived reproduction of the ICSFlux search
method to three frozen \name groundings. The study treats structural grounding
and dynamic realizability as separate evidentiary questions.

\section{Conclusion}
\label{sec:conclusion}

This paper examined when threat knowledge documented for one industrial
control system can be meaningfully evaluated on another. \name addresses this
question by transforming source evidence into a provenance-linked threat
abstraction and checking that abstraction against a target contract. Once the
abstraction, vocabulary, schema version, and contract are fixed, the
grounding checks are deterministic. An accepted result can be materialized as
a validation slice that records the mapped roles, signals, dependencies,
timing assumptions, and provenance. Slice adequacy, the dynamic reachability
of the effect, and the outcome of a downstream analysis remain separate
questions.

The grounding matrix evaluates 78 continuous-process abstractions under the
implemented target-contract filters. A comparison on a separately pinned
snapshot keeps candidate coverage distinct from five-check acceptance, and a
bounded audit, in which three judgment streams disagreed substantially,
indicates that returned candidates still require evidence-based review. The
most detailed executable evidence remains the water-domain
SWaT$\rightarrow$WT and WADI$\rightarrow$WD studies. Nine bounded-support
cases on Hydro and GRFICS chemical-process targets extend controlled slice
execution to additional target families; these are research-environment perturbations, not independently
captured cross-domain attacks. The study using the released upstream GeCo
implementation demonstrates compatibility for the evaluated slices. A
separate three-objective, one-target study shows that a grounding decision
can remain fixed while realizability changes with the evaluated plant model
and command surface.

These findings indicate that semantic correspondence alone does not establish
target-specific applicability under the implemented grounding criteria:
within the evaluated contracts and cases, the source interpretation had to
remain traceable to evidence, and candidate mappings had to satisfy explicit
target checks. Grounding results and downstream experimental results
should therefore be interpreted as answers to distinct questions, subject to
the limitations discussed in Section~\ref{sec:threats-validity}. Future work
will broaden the target-contract vocabulary, examine the individual grounding
criteria, compare stronger mapping approaches against independently
adjudicated mapping labels, and extend detailed executable evaluation to
additional target families and control architectures.

\begin{acks}
This material is based upon work supported in part by the U.S. National Science Foundation under Award Nos. 2425711 (FMitF) and 2330066 (SPHERE). Any opinions, findings, conclusions, or recommendations expressed in this material are those of the authors and do not necessarily reflect the views of the U.S. National Science Foundation.

Generative AI tools were used to assist with manuscript revision, language refinement, and figure preparation. LLMs were also explored for extracting structured intermediate representations (IRs) from ICS documentation, including threat semantics and
supporting evidence. These outputs were treated as provisional and were not used as the basis for the reported grounding results, which rely on structured or curated extraction paths. All AI-assisted content was reviewed by the authors, who retain full responsibility for the methodology, results, and conclusions.
\end{acks}

\clearpage
\appendix

\section{Source Evidence and Extraction Details}
\label{app:source-evidence}

Table~\ref{tab:app-source-evidence} provides additional source-evidence and
extraction details complementing the summary in
Table~\ref{tab:source-semantics}. It reports source- and domain-level evidence
and provenance metadata associated with the source abstractions used in the
evaluation. The evidence representation is informed
by OTThreat~\cite{paul2023towards}, while the evidence ledger and its
associated records are implemented as part of \name.

\begin{table}[!htbp]
\centering
\small
\caption{Source-evidence and extraction ledger used in the evaluation,
including provenance metadata.}
\label{tab:app-source-evidence}
%
\footnotesize
\begin{tabularx}{\columnwidth}{@{}p{0.27\columnwidth}X@{}}
\toprule
\textbf{Evidence object} & \textbf{Appendix detail} \\
\midrule

\textbf{Structured extraction} &
\textbf{Result:} 83 threats from 12 documents; 55 high-confidence, 18 medium-confidence, and 10 low-confidence. The source set includes SWaT~\cite{mathur2016swat} (51), WADI~\cite{ahmed2017wadi} (17), OilTreatment (10), and Fischertechnik (5). 
\textbf{Claim boundary:} these are coverage and confidence counts for the paper-facing corpus, not a full extraction precision/recall evaluation. \\[2pt]

\textbf{WADI~\cite{ahmed2017wadi} deepening} &
\textbf{Result:} naive PDF parsing yields 2 rows; \name recovers 15 A1 threats and 14 A2 attack windows, a 7.5$\times$ gain in attack-level threat records. 
\textbf{Claim boundary:} A1 seeds extraction, A2 contributes timing/observability windows, and A3 provides clean context; this is source-specific extraction, not a general PDF-table benchmark. \\[2pt]

\textbf{CISS~\cite{itrust2019ciss} semantics} &
\textbf{Result:} 187 parsed rows, 73 process-relevant rows, and 20 high-confidence derived threats from CISS~\cite{itrust2019ciss}. 
\textbf{Claim boundary:} CISS contributes source semantics and observability support; we do not treat it as perturbation replay. \\[2pt]

\textbf{\name evidence ledger} &
\textbf{Result:} 333 curated threats, 1,947 field-level evidence spans, and 100\% core-field coverage.
\textbf{Claim boundary:} the ledger is an artifact of \name, aligned with OTThreat~\cite{paul2023towards}; it supports provenance and auditability, not the grounding algorithm itself. \\

\bottomrule
\end{tabularx}

\end{table}

\section{Water-Domain Validation-Slice Details}
\label{app:water-slice-details}

Table~\ref{tab:app-water-slices} expands
Table~\ref{tab:flagship-slices} with adequacy, perturbation, and
consumer-outcome metadata for the water-domain validation slices. Raw consumer
responses are reported separately from nominal-subtracted outcomes because
some violations observed during the perturbation window are also present in
the corresponding nominal traces. Nominal subtraction is therefore used to
distinguish responses that remain after accounting for the nominal baseline
from nominal-confounded outcomes and to identify cases near the evaluated
consumer's decision boundary.

\begin{table}[!htbp]
\centering
\small
\caption{Detailed water-domain validation-slice evidence, including slice
adequacy, controlled-perturbation metadata, raw consumer response,
nominal-subtracted outcome, and interpretation. Target signals are defined by
the corresponding SPHERE target contracts.}
\label{tab:app-water-slices}
\footnotesize
\begin{tabularx}{\columnwidth}{@{}p{0.30\columnwidth}X@{}}
\toprule
\textbf{Grounding case} & \textbf{Detailed outcome} \\
\midrule

\textbf{SWaT$\to$WT: tank-level high spoof} &
\textbf{Adequacy:} target signals declared in the WT contract; 100\% source/target/dependency coverage; timing evaluable.
\textbf{Outcome:} nominal-subtracted delta yields 3 violations: 2 rate-of-change and 1 mass-balance.
\textbf{Interpretation:} clean WT invariant case. \\[2pt]

\textbf{SWaT$\to$WT: pump-flow low spoof} &
\textbf{Adequacy:} target signals declared in the WT contract; 100\% source/target/dependency coverage; timing evaluable.
\textbf{Outcome:} raw analysis yields 9 correlation violations, but nominal-subtracted delta yields 0.
\textbf{Interpretation:} nominal-confounded; not counted as a clean invariant success. \\[2pt]

\textbf{WADI$\to$WD: supply-flow low spoof} &
\textbf{Adequacy:} target signals declared in the WD contract; 100\% source/target/dependency coverage; partial verdict.
\textbf{Outcome:} no baseline invariant violations; first firing occurs at $-64$ with shallow range/correlation evidence.
\textbf{Interpretation:} near-threshold boundary case. \\[2pt]

\textbf{WADI$\to$WD: supply-flow high idle} &
\textbf{Adequacy:} target signals declared in the WD contract; 100\% source/target/dependency coverage; strong verdict.
\textbf{Outcome:} 11 range violations.
\textbf{Interpretation:} clean WD flow/pressure case. \\[2pt]

\textbf{WADI$\to$WD: NaOCl level step} &
\textbf{Adequacy:} target signals declared in the WD contract; 100\% source/target/dependency coverage; strong verdict.
\textbf{Outcome:} 3 mass-balance/rate-of-change violations.
\textbf{Interpretation:} clean WD dosing case; outside the GeCo-style predictive surface. \\

\bottomrule
\end{tabularx}

\end{table}

\clearpage
\section{Bounded-Support Validation-Slice Details (Hydro/GRFICS)}
\label{app:hydro-grfics-details}

Table~\ref{tab:app-hydro-grfics} provides the detailed adequacy and
consumer-response metadata for the nine bounded-support cases summarized in
Table~\ref{tab:hydro-grfics-slices}. All nine satisfy the declared adequacy
requirements for their respective studies. Nominal-subtracted baselines are
unavailable for these cases, so the observed rule responses are not assigned
the \emph{clean} consumer-outcome classification used for flagship cases with
corresponding nominal evidence. The earliest recorded rule violation occurs
at the perturbation onset (sample 20) in all nine cases. This timing is
reported as execution metadata supporting slice executability and should not
be interpreted as a substitute for nominal-subtracted evidence or as evidence
of independently executed attacks.

\begin{table}[!htbp]
\centering
\footnotesize
\caption{Adequacy and consumer-response metadata for the nine Hydro/GRFICS
bounded-support cases. ``Earliest'' denotes the sample index of the first
recorded rule violation; the controlled perturbation begins at sample 20 for
every case. Observed rule responses characterize bounded consumer behavior
without a nominal-subtracted baseline.}
\label{tab:app-hydro-grfics}
\begin{tabularx}{\linewidth}
{@{}p{0.22\linewidth}p{0.14\linewidth}p{0.05\linewidth}p{0.05\linewidth}X@{}}
\toprule
\textbf{Case} &
\textbf{Expected rule types} &
\textbf{Viol.} &
\textbf{Earl.} &
\textbf{Observed rule types (expected hit / total)} \\
\midrule

swat\_hydro\_uc1 / hydro\_reservoir\_level\_high\_spoof &
range; rate\_of\_change &
23 & 20 &
range; rate\_of\_change (2/2) \\

swat\_hydro\_uc1 / hydro\_speed\_pct\_high\_spoof &
range; rate\_of\_change &
23 & 20 &
range; rate\_of\_change (2/2) \\

swat\_hydro\_uc1 / hydro\_flow\_low\_spoof &
range; correlation; causality &
23 & 20 &
range; rate\_of\_change (1/3 expected; additional rate\_of\_change) \\

wadi\_grfics\_uc1 / grfics\_tank\_pressure\_high\_spoof &
range; rate\_of\_change &
2 & 20 &
rate\_of\_change (1/2) \\

wadi\_grfics\_uc1 / grfics\_tank\_level\_high\_spoof &
range; rate\_of\_change &
3 & 20 &
causality; rate\_of\_change (1/2 expected; additional causality) \\

wadi\_grfics\_uc1 / grfics\_feed1\_flow\_low\_spoof &
range; correlation; causality &
25 & 20 &
causality; range; rate\_of\_change
(2/3 expected; additional rate\_of\_change) \\

epic\_hydro\_uc0 / hydro\_speed\_pct\_high\_spoof &
range; rate\_of\_change &
23 & 20 &
range; rate\_of\_change (2/2) \\

epic\_hydro\_uc0 / hydro\_power\_mw\_low\_spoof &
range; correlation &
22 & 20 &
correlation; range (2/2) \\

epic\_hydro\_uc0 / hydro\_flow\_low\_spoof &
range; correlation; causality &
23 & 20 &
range; rate\_of\_change
(1/3 expected; additional rate\_of\_change) \\

\bottomrule
\end{tabularx}
\end{table}

\clearpage
\section{Bounded Consumer Implementations}
\label{app:consumer-implementations}

Table~\ref{tab:app-consumers} documents the implementation scope of each
bounded consumer lane over the common \name validation-slice representation.
The GeCo-, SAIN-, and SCAPHY-style lanes are \name-native implementations of
selected predictive, state-aware, and phase-aware analysis patterns motivated
by the corresponding published systems. They are used to evaluate whether the
validation-slice representation provides the signals and contextual
information required by these analysis styles. Compatibility with the released
upstream GeCo implementation is evaluated separately in
Section~\ref{sec:eval-geco-real}.

\begin{table}[!htbp]
\centering
\small
\caption{Bounded consumer implementations over the common validation-slice
representation. The ``-style'' lanes are \name-native implementations of
selected analysis patterns motivated by the corresponding published systems
and are evaluated as bounded consumers rather than reproductions of those
systems.}
\label{tab:app-consumers}
\footnotesize
\begin{tabularx}{\columnwidth}
{@{}p{0.30\columnwidth}X@{}}
\toprule
\textbf{Consumer family} & \textbf{Bounded implementation details} \\
\midrule

\textbf{Invariant checks} &
\textbf{Slice inputs:} signal ranges, correlations, mass-balance relationships, and rate-of-change constraints.
\textbf{Calibration:} thresholds from nominal traces.
\textbf{Scoring:} violation count and nominal-subtracted delta.
\textbf{Boundary:} native \name consumer and primary validation lane. \\[2pt]

\textbf{GeCo-style predictive~\cite{wolsing2025geco}} &
\textbf{Slice inputs:} time-series windows and signal correlations.
\textbf{Calibration:} nominal-trace training and threshold calibration.
\textbf{Scoring:} hit/miss, nominal FPR, and F1.
\textbf{Boundary:} bounded predictive lane inspired by GeCo; not upstream code reuse or faithful reproduction. \\[2pt]

\textbf{SAIN-style state-aware~\cite{abbas2024sain}} &
\textbf{Slice inputs:} target signals and coarse discrete process state.
\textbf{Calibration:} state-conditioned thresholds from nominal traces.
\textbf{Scoring:} hit/miss, nominal FPR, and F1.
\textbf{Boundary:} bounded state-aware lane; not full SAIN reproduction. \\[2pt]

\textbf{SCAPHY-style phase-aware~\cite{ike2022scaphy}} &
\textbf{Slice inputs:} target signals, phase labels, and operating context.
\textbf{Calibration:} phase-segmented nominal calibration.
\textbf{Scoring:} hit/miss, nominal FPR, and F1.
\textbf{Boundary:} bounded phase/context-aware lane; not full SCAPHY reproduction. \\

\bottomrule
\end{tabularx}
\end{table}

\clearpage
\section{Additional Downstream Adapter Evidence}
\label{app:downstream-adapters}

Table~\ref{tab:app-downstream-adapters} summarizes \name-authored prototype
adapters for post-analysis classification and provenance-oriented workflows.
The adapters use analysis patterns and output structures motivated by SCADMAN
and ICSTracker but do not execute the published systems. Their measurements
therefore characterize the \name-authored prototypes rather than the
corresponding published systems. This evidence is distinct from the upstream
GeCo study in Section~\ref{sec:eval-geco-real}, which executes the released
GeCo pipeline over \name-grounded validation slices.

The reviewed \texttt{SCADMAN-Light} artifact was not treated as an
implementation of the published SCADMAN analysis in this evaluation. Within
the reviewed artifact, the available implementation applies a weighted
attack-signature classifier with a fixed decision threshold rather than
exposing the control-flow-attestation mechanism considered in our
SCADMAN-motivated adapter.

\noindent\textbf{Post-analysis classification adapter.}
The prototype consumes detector windows, evaluates target-grounded temporal
predicates, and emits typed events with robustness and provenance metadata.
Under the evaluated operating point of one false alarm per hour on benign
segments (1~FA/hr), applied to both configurations, adding the \name-derived
semantic classification stage changes precision from 0.375 to 1.000 and F1 from 0.545
to 0.800, while recall changes from 1.000 to 0.667. These measurements
characterize the evaluated \name prototype and operating point.

\noindent\textbf{Provenance-workflow adapter.}
The prototype consumes grounded mappings, eligibility outcomes, projected
edges, impact chains, and lineage records and instantiates rule templates over
eligible mappings. In the evaluated workflow, all 17 templates are
instantiable and produce 2,074 typed mappings. The corresponding rerun
agreement measurement is 0.95 under the evaluated configuration. These
measurements characterize the \name prototype workflow and do not constitute
an execution or evaluation of the published ICSTracker system.

\begin{table}[!htbp]
\centering
\footnotesize
\caption{Prototype downstream-adapter evidence. Measurements characterize the
\name-authored adapters under the evaluated configurations; SCADMAN and
ICSTracker motivate the corresponding analysis patterns but are not executed
as baselines. FA/hr denotes false alarms per hour on benign segments, used
here as a fixed operating point; P/R/F1 denote precision, recall, and F1
score.}
\label{tab:app-downstream-adapters}

\footnotesize
\begin{tabularx}{\columnwidth}{@{}p{0.30\columnwidth}X@{}}
\toprule
\textbf{Prototype adapter} &
\textbf{Verified prototype evidence and claim boundary} \\
\midrule

\textbf{Post-analysis classification/typing} &
\textbf{Result:} at 1~FA/hr, baseline P/R/F1 was
0.375/1.000/0.545; adding \name semantics yielded
1.000/0.667/0.800, corresponding to a precision gain of +0.625
and F1 gain of +0.255.
\textbf{What it exercises:} post-analysis typing over detector windows
using grounded temporal predicates and provenance.
\textbf{Boundary:} these measurements characterize an earlier
\name prototype adapter only. SCADMAN is cited as motivating literature
and was not independently reproduced here; the evaluated prototype does
not implement SCADMAN's control-flow-attestation mechanism and therefore
does not measure SCADMAN performance or mechanism fidelity. \\[2pt]

\textbf{Provenance-workflow adapter} &
\textbf{Result:} 17/17 rule templates instantiated, 2,074 typed
physics-consistent mappings, and 0.95 reproducibility across reruns.
\textbf{What it exercises:} rule expansion, provenance-pack generation,
eligibility checks, and typed path export.
\textbf{Boundary:} these measurements characterize the prototype
provenance workflow only. ICSTracker is cited as motivating literature;
because implementation fidelity has not been independently established,
these results do not constitute an ICSTracker reproduction or performance
evaluation. \\

\bottomrule
\end{tabularx}
\end{table}

\clearpage
\section{Extension Evidence Ledger}
\label{app:extension-evidence}

Table~\ref{tab:app-extension-evidence} provides additional details for the
extension studies summarized in Section~\ref{sec:eval-extensions}. These
studies exercise evidence surfaces distinct from the primary
continuous-process evaluation, including source-semantic breadth,
manufacturing/program-artifact grounding, controller-code enrichment,
temporal guardrails, protocol evidence, and event/timing obligations.
Accordingly, their results use study-specific denominators and are reported
separately from the continuous-process grounding percentages and primary
WT/WD validation-slice studies.

\begin{table*}[!htbp]
\centering
\small
\caption{Extension evidence ledger. Each row identifies the evaluated
extension, its principal result, the \name capability exercised, and the
evidentiary scope of the result.}
\label{tab:app-extension-evidence}
%

\begingroup
\setlength{\tabcolsep}{2pt}
\renewcommand{\arraystretch}{1.12}
\scriptsize
\begin{tabular}{@{}>{\raggedright\arraybackslash}p{0.14\linewidth}>{\raggedright\arraybackslash}p{0.22\linewidth}>{\raggedright\arraybackslash}p{0.18\linewidth}>{\raggedright\arraybackslash}p{0.22\linewidth}>{\raggedright\arraybackslash}p{0.14\linewidth}@{}}
\toprule
\textbf{Extension} & \textbf{Key Result} & \textbf{What It Exercises} & \textbf{Why It Is Bounded} & \textbf{Paper Loc.} \\
\midrule

CISS observability
& 20 high-confidence threats; 13/20 full-WT grounded; avg score 2.45
& Source semantics breadth; slice-minimality
& Source semantics only, not perturbation replay
& \S\ref{sec:eval-source} \\
\addlinespace[2pt]

TXT$\to$S7 Fischertechnik
& 5/5 threats grounded; 102 tags; 26/28 components
& Manufacturing/program-artifact grounding
& Separate manufacturing denominator; not continuous-process matrix
& \S\ref{sec:eval-extensions}, App.~\ref{app:extension-evidence} \\
\addlinespace[2pt]

CrossPLC control study
& MiniSWaT/SPHERE-WT: 6/6 role families; direct matrix delta 0
& Controller-code enrichment; provenance
& Optional enrichment; does not improve grounding percentages
& \S\ref{sec:eval-extensions}, App.~\ref{app:extension-evidence} \\
\addlinespace[2pt]

STL/RTAMT guardrail
& WADI/WD: $-63$ accepted, $-64$ rejected; $+73$ accepted, $+74$ rejected
& Formal slice-level temporal guardrail
& One bounded WADI$\to$WD branch; not full verification
& \S\ref{sec:eval-extensions}, App.~\ref{app:extension-evidence} \\
\addlinespace[2pt]

EPIC/Hydro scenarios
& 6 direct + 2 bridge scenarios; 0.8 semantic coverage; 3/3 strong support
& Power/hydro domain breadth
& Scenario support only; not EPIC attack-trace replay
& \S\ref{sec:eval-extensions}, App.~\ref{app:extension-evidence} \\
\addlinespace[2pt]

PowerDuck/\newline GOOSE
& 16 IPAL files; 1,133,589 packets; flooding/insertion/replay/suppression
& Protocol-level evidence complement
& Protocol complement to EPIC; not target replay
& \S\ref{sec:eval-extensions}, App.~\ref{app:extension-evidence} \\
\addlinespace[2pt]

VetPLC-style obligations
& 10 scenarios; 2 ST programs; 1 TON timer; 6 IF conditions; 12 variables
& Event/timing obligation extraction
& Obligation artifacts only; no VetPLC engine reproduction
& \S\ref{sec:eval-extensions}, App.~\ref{app:extension-evidence} \\

\bottomrule
\end{tabular}
\endgroup

\end{table*}

\begin{figure}[!htbp]
\centering
\includegraphics[width=0.85\linewidth]
{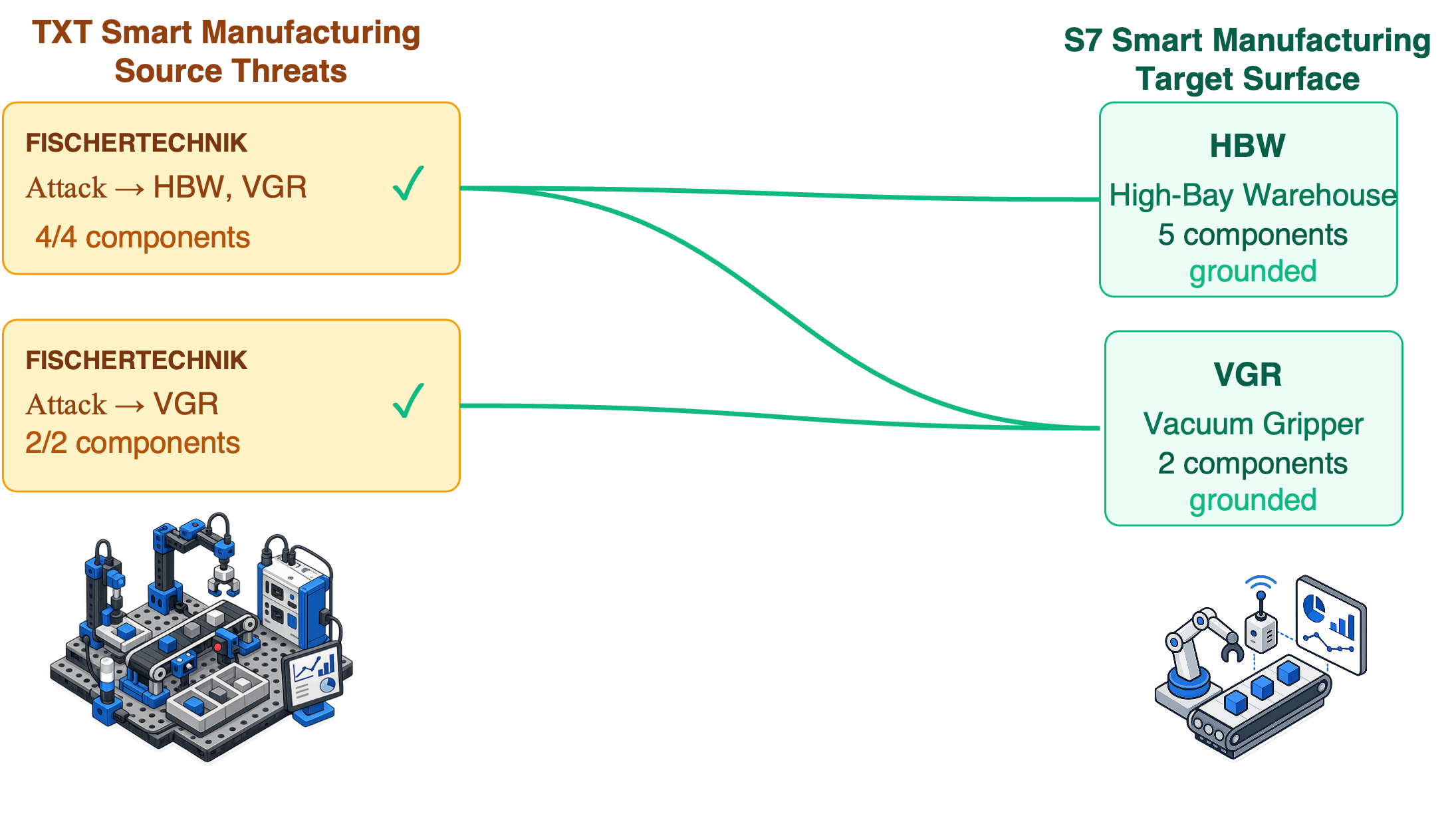}
\caption{TXT$\rightarrow$S7 manufacturing/program-artifact grounding.
Fischertechnik/TXT source abstractions are evaluated against a Siemens S7
target surface using the available program, tag, and state evidence. The
study is reported under a manufacturing/program-artifact denominator separate
from the continuous-process grounding matrix in
Figure~\ref{fig:grounding-matrix}.}
\label{fig:txt-s7}
\end{figure}

Figure~\ref{fig:txt-s7} illustrates the TXT$\rightarrow$S7
manufacturing/program-artifact grounding study. This study uses a
manufacturing denominator distinct from the 78-threat continuous-process
grounding corpus represented in Figure~\ref{fig:grounding-matrix}.

\clearpage
\section{Pinned-Snapshot Candidate Comparison and Bounded Assignment Audit}
\label{app:mapping-audit}

This appendix details the candidate comparison and audit summarized in
Section~\ref{sec:eval-grounding}. Both use a separately pinned input snapshot,
separate from the run reported in Figure~\ref{fig:grounding-matrix}. All reported
quantities are aggregate counts.

\subsection{Snapshot and Comparison Population}

The snapshot manifest pins the \name source artifact, grounder, grounding
configuration, baseline module, and runner, together with their SHA-256
content digests, as frozen evaluation artifacts. It pins the four target tag
contracts and the four defense-rule files to fixed versions of the SPHERE
use-case repository and the CPS enclave model. The grounding configuration's
chemical-process contract path was reconciled explicitly to the
Tennessee-Eastman contract, which had been renamed without any content
change. The runner produced \name results for
all 332 source--target pairs (83 sources $\times$ four targets). The 312
pairs formed by the 78 continuous-process sources enter the comparison;
the five Fischertechnik records keep their separate denominator. Compared
with the per-pair outputs underlying Figure~\ref{fig:grounding-matrix}, all 332 pair identities matched, with no
differences in acceptance, ordered mapped tags, ordered applicable rules, or
any of the five eligibility dimensions. This shows output-level agreement
only.

Table~\ref{tab:app-new-snapshot-cells} reports each source-domain $\times$
target cell. The literal name baseline returned zero candidate-bearing pairs
and zero mapped components in every cell and is therefore omitted from the
per-cell columns. Candidate coverage and five-check acceptance are different
measures: a candidate-bearing pair may still fail an eligibility check, and
neither quantity is mapping correctness. Both are assessed at the pair level.

\name returned candidates for 16 more pairs than the role-only baseline (222
versus 206). The Hydro column accounts for $+17$ of this difference (SWaT
$+6$, WADI $+8$, OilTreatment $+3$), and WADI$\rightarrow$WT for $-1$. The
remaining eight cells are equal. \name's candidate construction also uses
configured signal aliases and target role preferences that the baselines do
not receive, so the difference reflects this additional target-specific
configuration as well as the matching procedure. The comparison therefore
does not isolate an algorithmic improvement.

\begin{table}[!htbp]
\centering
\footnotesize
\caption{Pinned-snapshot candidate comparison by source domain and target.
``Comp.'' counts source components once per target. ``\name acc.'' is
five-check acceptance. ``R/X'' gives role-only baseline and \name
candidate-construction counts. ``R$=$X'' counts pairs with identical
nonempty assignments, and ``Both abst.'' counts pairs where both methods
returned nothing. The literal name baseline is zero in every cell. All
entries are counts, not success or correctness rates.}
\label{tab:app-new-snapshot-cells}
\begingroup
\setlength{\tabcolsep}{4pt}
\begin{tabular}{@{}lrrrrrrr@{}}
\toprule
\textbf{Source $\rightarrow$ target} &
\textbf{Pairs} &
\textbf{Comp.} &
\textbf{\name acc.} &
\textbf{Cand.\ pairs R/X} &
\textbf{Mapped comp.\ R/X} &
\textbf{R$=$X} &
\textbf{Both abst.} \\
\midrule
SWaT $\rightarrow$ WT & 51 & 127 & 38 & 38/38 & 98/98 & 0 & 13 \\
SWaT $\rightarrow$ WD & 51 & 127 & 37 & 37/37 & 94/94 & 1 & 14 \\
SWaT $\rightarrow$ Hydro & 51 & 127 & 30 & 25/31 & 33/48 & 8 & 20 \\
SWaT $\rightarrow$ GRFICS/TE & 51 & 127 & 20 & 31/31 & 48/48 & 20 & 20 \\
WADI $\rightarrow$ WT & 17 & 26 & 15 & 16/15 & 25/23 & 0 & 1 \\
WADI $\rightarrow$ WD & 17 & 26 & 14 & 14/14 & 21/21 & 2 & 3 \\
WADI $\rightarrow$ Hydro & 17 & 26 & 12 & 5/13 & 6/20 & 3 & 4 \\
WADI $\rightarrow$ GRFICS/TE & 17 & 26 & 11 & 13/13 & 20/20 & 5 & 4 \\
OilTreatment $\rightarrow$ WT & 10 & 13 & 8 & 8/8 & 10/10 & 0 & 2 \\
OilTreatment $\rightarrow$ WD & 10 & 13 & 8 & 8/8 & 10/10 & 1 & 2 \\
OilTreatment $\rightarrow$ Hydro & 10 & 13 & 7 & 4/7 & 4/8 & 1 & 3 \\
OilTreatment $\rightarrow$ GRFICS/TE & 10 & 13 & 5 & 7/7 & 8/8 & 3 & 3 \\
\midrule
\textbf{Total} & \textbf{312} & \textbf{664} & \textbf{205} &
\textbf{206/222} & \textbf{377/408} & \textbf{44} & \textbf{89} \\
\bottomrule
\end{tabular}
\endgroup

\end{table}

\subsection{Audit Selection Rule}

The audit selection rule was frozen before any judgment was collected. The
executed selector is the \texttt{select\_cases} function of the frozen \name
audit module used to prepare the cards; its selection logic is unchanged in
later versions. Within each of the 12 source-domain $\times$ target cells, it considers only
pairs for which at least one method returned an assignment. Eligible pairs are ranked
by the SHA-256 digest of a fixed protocol label, a fixed purpose label, and
the pair identifier. The frozen protocol specified one
method-disagreement pair and one method-agreement pair per cell, with
hash-ranked fallback. Here, \emph{method agreement} means that the
candidate methods returned identical assignments; it is unrelated to
agreement among the judgment streams reported in
Table~\ref{tab:app-audit-agreement}. The implementation tests agreement across
all three methods, including the literal baseline. Because the literal
baseline returned no assignment in any pair, every eligible pair counts as a
three-method disagreement, and no eligible agreement pair exists. The
intended agreement/disagreement enrichment therefore did not take effect.
In each cell, the rule selected the two highest-ranked eligible pairs: the
first as the disagreement pick and the second through fallback. This yields
24 pairs. The selection does not target pairs on which the role-only
baseline and \name agree or disagree. Its only systematic constraints are
the exclusion of the 89 pairs on which both remaining methods abstained and
the fixed allocation of two pairs to every cell, regardless of cell size (10
to 51 pairs).

\subsection{Evidence Cards}

Each selected pair became a method-blind card. A card contains the structured
source semantics (effect family; source components with identifier, type,
role, and process stage; physical consequence; and mechanism tags) and the
target-contract metadata (name, type, direction, role, group, units, and
description) of the tags appearing in the pair's returned assignments. The
card lists the union of all methods' returned assignments in a hash-shuffled
order without method identities. Cards omit raw licensed source documents,
simulator and target implementation internals, and some evidence that may
inform \name mappings, such as configured aliases and role preferences. An
\emph{insufficiently evidenced} verdict on a card is therefore not by itself
evidence of a mapping error. The 24 cards contain 116 distinct returned
assignments: 61 from the role-only baseline and 65 from \name, of which 10
were returned by both; the literal baseline returned none.

Grounding eligibility is assessed per source--target pair, whereas the audit
judges individual returned assignments. The assignments are unevenly
distributed across the 24 pairs. One pair contributes 42 of the 116
assignments, and the two SWaT$\rightarrow$WD pairs together contribute 47.
The pooled totals in Tables~\ref{tab:app-audit-verdicts}
and~\ref{tab:app-audit-agreement} are therefore assignment-level counts
clustered within 24 pairs, not judgments of 116 independent cases. They weight
each pair by its number of assignments, so a single pair can dominate them.

\subsection{Judgment Streams and Custody}

Three streams labeled every assignment as \emph{supported},
\emph{insufficiently evidenced}, or \emph{unsupported}. The sealing step
accepts a stream only if every assignment has one of these verdicts and a
non-empty reason. One stream is the \emph{human reviewer}, one of the paper's
authors. It is a human judgment stream, not an
independent external expert and not a ground-truth oracle. The other two are
distinct OpenAI
Codex models (\texttt{gpt-6-sol} and \texttt{gpt-6-luna}, as recorded in
their sealed streams), which share a provider and are neither independent of
each other nor human experts. The language models served only as audit
judgment instruments in this study. Their verdicts are reported as data, not
as ground truth, and they did not produce the grounding or baseline results.

The human reviewer used a local, method-blind review page and written packet
instructions. These directed the reviewer to judge each distinct returned
assignment rather than each case, using only the structured source semantics
and target-tag metadata on the card. The instructions stated that the review
concerns returned tag assignments, not the original source document or an
attack-execution outcome, and that correctness should not be inferred from a
method name. The review page also displayed brief interface guidance to
choose \emph{insufficiently evidenced} when the card cannot establish support
either way. This guidance was a usage note on the review page, not a
prespecified operational rubric.

The two model streams were run with reasoning effort set to high. Their
inputs were the 24 blinded card files and the manifest. The models had no
access to the comparison input, the method key, case-level results in the
\name repository, each other's outputs, or licensed source data. Their task
instructions required them to preserve card and assignment identifiers and
not to infer attack execution or detector outcomes. The instructions also
described the model streams as exploratory rather than independent oracles.

For all three streams, the recorded instructions provide the three verdict
labels, restrict judgments to the blinded card evidence, and require a brief
evidence-based reason.

The model-task wording was recovered retrospectively from retained agent
context. The archived task payload is encrypted, and no separately preserved
plaintext prompt artifact was identified.

Each stream was sealed with a digest bound to the frozen cards and manifest.
After all three seals existed, a \emph{reveal} step validated them against
the frozen comparison input and wrote the private method key. A separate
\emph{summarize} step validated the seals and key again and wrote the
aggregate counts reported here.

The human reviewer's packet contained the blind cards and, in a separate
folder, the frozen comparison input and the two model seals. Both folders
were delivered together. The human reviewer attests to sealing the
human-review stream before opening the second folder, so this separation is
procedural self-custody, not an independently enforced blind. The sealed
human-review stream
was independently validated against the frozen cards, manifest, and
comparison. Repeated reveal and summary runs reproduced the aggregate counts
exactly after Windows line endings were normalized. Case-level judgments, reasons, cards, and the method key
are held privately.

\subsection{Aggregate Verdicts and Agreement}

Table~\ref{tab:app-audit-verdicts} gives verdict totals over all 116 distinct
assignments, and Table~\ref{tab:app-audit-agreement} gives the number of
assignments on which streams gave identical verdicts. Verdicts are not
broken down by method here, and the counts should not be used to rank
methods.

\begin{table}[!htbp]
\centering
\footnotesize
\caption{Verdict totals over the 116 distinct returned assignments in the
24-pair stratified audit selection. These are card-limited judgments, not mapping
accuracy; no stream is ground truth.}
\label{tab:app-audit-verdicts}
\begin{tabular}{@{}lrrr@{}}
\toprule
\textbf{Judgment stream} & \textbf{Supported} &
\textbf{Insufficiently evidenced} & \textbf{Unsupported} \\
\midrule
Human reviewer & 77 & 37 & 2 \\
Model A (\texttt{gpt-6-sol}) & 67 & 15 & 34 \\
Model B (\texttt{gpt-6-luna}) & 41 & 47 & 28 \\
\bottomrule
\end{tabular}
\end{table}

\begin{table}[!htbp]
\centering
\footnotesize
\caption{Identical verdicts between judgment streams over the 116 distinct
returned assignments.}
\label{tab:app-audit-agreement}
\begin{tabular}{@{}lr@{}}
\toprule
\textbf{Stream comparison} & \textbf{Identical verdicts (/116)} \\
\midrule
Human reviewer vs.\ Model A & 74 \\
Human reviewer vs.\ Model B & 47 \\
Model A vs.\ Model B & 66 \\
All three streams & 36 \\
\bottomrule
\end{tabular}
\end{table}

\subsection{Interpretation and Limitations}

The audit shows substantial disagreement among three card-limited judgment
streams about the same returned assignments: pairwise identical verdicts
range from 47 to 74 of 116, and the streams' unsupported totals range from 2
to 34. We use this result only as diagnostic caution:
returned candidates, including those passing \name's eligibility checks,
still require evidence-based review. The audit does not estimate mapping
accuracy, does not compare methods statistically, and does not constitute
a user study. Its limitations are the stratified, non-proportional
selection that excludes both-abstain pairs; the concentration of assignments
in a few pairs; two model streams from the same provider; the absence of written
operational definitions for the three verdicts; model-task instructions that
were recovered retrospectively; procedural rather than
enforced
blinding; and the absence of adjudication or an independent oracle. Cards
omit raw source documents and some \name evidence. This omission may
contribute to verdicts on \name assignments, but the audit does not
establish that it explains the observed disagreement. The earlier review
of all 83 paper-facing source records
(Section~\ref{sec:eval-source}) checked source locations and excerpts only.

\clearpage

\bibliographystyle{ACM-Reference-Format}
\bibliography{Sources/references}

\end{document}